\documentclass[article,pdftex,shortnames,nojss]{jss}\usepackage[]{graphicx}\usepackage[]{color}
\makeatletter
\def\maxwidth{ %
  \ifdim\Gin@nat@width>\linewidth
    \linewidth
  \else
    \Gin@nat@width
  \fi
}
\makeatother

\definecolor{fgcolor}{rgb}{0.345, 0.345, 0.345}

\usepackage{framed}
\makeatletter
\newenvironment{kframe}{%
 \def\at@end@of@kframe{}%
 \ifinner\ifhmode%
  \def\at@end@of@kframe{\end{minipage}}%
  \begin{minipage}{\columnwidth}%
 \fi\fi%
 \def\FrameCommand##1{\hskip\@totalleftmargin \hskip-\fboxsep
 \colorbox{shadecolor}{##1}\hskip-\fboxsep
     \hskip-\linewidth \hskip-\@totalleftmargin \hskip\columnwidth}%
 \MakeFramed {\advance\hsize-\width
   \@totalleftmargin\z@ \linewidth\hsize
   \@setminipage}}%
 {\par\unskip\endMakeFramed%
 \at@end@of@kframe}
\makeatother

\definecolor{shadecolor}{rgb}{.97, .97, .97}
\definecolor{messagecolor}{rgb}{0, 0, 0}
\definecolor{warningcolor}{rgb}{1, 0, 1}
\definecolor{errorcolor}{rgb}{1, 0, 0}
\newenvironment{knitrout}{}{} 

\usepackage{alltt}
\usepackage[ruled,noline,linesnumbered]{algorithm2e}
\SetKwFor{For}{for}{do}{end}
\SetKwFor{While}{while}{do}{end}
\SetKwInput{KwIn}{input}
\SetKwInput{KwOut}{output}
\SetKwInput{KwCplx}{complexity}
\SetKwBlock{Begin}{procedure}{}
\DontPrintSemicolon

\usepackage{paralist}
\usepackage{amsmath}
\usepackage{amssymb}
\usepackage{color}
\usepackage{graphicx}
\usepackage[all]{hypcap}

\usepackage[sort&compress]{cleveref}
\newcommand{\crefrangeconjunction}{--}
\crefname{figure}{Fig.}{Figs.}
\Crefname{figure}{Fig.}{Figs.}
\crefname{table}{Table}{Tables}
\Crefname{table}{Table}{Tables}
\crefname{equation}{Eq.}{Eqs.}
\Crefname{equation}{Eq.}{Eqs.}
\creflabelformat{equation}{#2#1#3}
\crefname{appendix}{Appendix}{Appendices}
\Crefname{appendix}{Appendix}{Appendices}
\crefname{algorithm}{Algorithm}{Algorithms}
\Crefname{algorithm}{Algorithm}{Algorithms}
\crefname{section}{Section}{Sections}
\Crefname{section}{Section}{Sections}
\crefname{AlgoLine}{line}{lines}
\Crefname{AlgoLine}{Line}{Lines}

\newcommand\featureFunc{\mathbb{S}}
\newcommand\loglik{\ell}
\newcommand\loglikMC{\hat\ell}
\newcommand\synloglik{\loglik_{\featureFunc}}
\newcommand\synloglikMC{\loglikMC_{\featureFunc}}
\newcommand\R{\mathbb{R}}
\newcommand\slot[1]{\textit{#1}}
\newcommand\class[1]{class `\textrm{#1}'}
\newcommand\prob[1]{\mathbb{P}\left[{#1}\right]}
\newcommand\expect[1]{\mathbb{E}\left[{#1}\right]}
\newcommand\var[1]{\mathrm{Var}\left[{#1}\right]}

\newcommand\given{{\,\vert\,}}

\newcommand\myequals{\hspace{0.5mm}{=}\hspace{0.5mm}}
\newcommand\myto{{\;:\;}}
\newcommand\seq[2]{{#1}\!:\!{#2}}
\newcommand\mydot{{\,\cdot\,}}
\newcommand\cp[2]{N_{\mathrm{#1}\mathrm{#2}}}
\newcommand\BirthDeath{\raisebox{-0.3ex}{\scalebox{1.5}{$\cdot$}}}
\newcommand\dlta[1]{\Delta{#1}}
\newcommand\giventh{{\hspace{0.5mm};\hspace{0.5mm}}}
\newcommand\normal{{\mathrm{Normal}}}
\newcommand\argequals{{\,=\,}}
\newcommand\lags{c}
\newcommand\maxlag{\overline{c}}

\newcommand\bigO[1]{\mathcal{O}\!\left({#1}\right)}

\author{Aaron A. King, Dao Nguyen, Edward L. Ionides\\University of Michigan\\ \today}
\title{Statistical Inference for Partially Observed Markov Processes via the \proglang{R} Package \pkg{pomp}}
\Plainauthor{Aaron A. King, Dao Nguyen, Edward L. Ionides} 
\Plaintitle{Statistical inference for partially observed Markov processes via the R package pomp} 
\Shorttitle{Partially observed Markov processes}

\Abstract{
Partially observed Markov process (POMP) models, also known as hidden Markov models or state space models, are ubiquitous tools for time series analysis.
The R package \pkg{pomp} provides a very flexible framework for Monte~Carlo statistical investigations using nonlinear, non-Gaussian POMP models.
A range of modern statistical methods for POMP models have been implemented in this framework including sequential Monte~Carlo, iterated filtering, particle Markov chain Monte~Carlo, approximate Bayesian computation, maximum synthetic likelihood estimation, nonlinear forecasting, and trajectory matching.
In this paper, we demonstrate the application of these methodologies using some simple toy problems.
We also illustrate the specification of more complex POMP models, using a nonlinear epidemiological model with a discrete population, seasonality, and extra-demographic stochasticity.
We discuss the specification of user-defined models and the development of additional methods within the programming environment provided by \pkg{pomp}.

\noindent
*This document is a version of a manuscript in press at the \emph{Journal of Statistical Software}.
It is provided under the \href{http://creativecommons.org/licenses/by/3.0/}{Creative Commons Attribution License}.
}
\Plainkeywords{Markov processes, hidden Markov model, state space model, stochastic dynamical system, maximum likelihood, plug-and-play, time series, mechanistic model, sequential Monte Carlo, R}
\Address{
  Aaron A. King\\
  Departments of Ecology \& Evolutionary Biology and Mathematics\\
  Center for the Study of Complex Systems\\
  University of Michigan\\
  48109 Michigan, USA\\
  E-mail: \email{kingaa@unich.edu}\\
  URL: \url{http://kinglab.eeb.lsa.umich.edu/}\\

  Dao Nguyen\\
  Department of Statistics\\
  University of Michigan\\
  48109 Michigan, USA\\
  E-mail: \email{nguyenxd@unich.edu}\\

  Edward Ionides\\
  Department of Statistics\\
  University of Michigan\\
  48109 Michigan, USA\\
  E-mail: \email{ionides@unich.edu}\\
  URL: \url{http://www.stat.lsa.umich.edu/~ionides}\\

}

\IfFileExists{upquote.sty}{\usepackage{upquote}}{}
\begin{document}

\section {Introduction}

A partially observed Markov process (POMP) model consists of incomplete and noisy measurements of a latent, unobserved Markov process.
The far-reaching applicability of this class of models has motivated much software development \citep{commandeur11}.
It has been a challenge to provide a software environment that can effectively handle broad classes of POMP models and take advantage of the wide range of statistical methodologies that have been proposed for such models.
The \pkg{pomp} software package \citep{pomp} differs from previous approaches by providing a general and abstract representation of a POMP model.
Therefore, algorithms implemented within \pkg{pomp} are necessarily applicable to arbitrary POMP models.
Moreover, models formulated with \pkg{pomp} can be analyzed using multiple methodologies in search of the most effective method, or combination of methods, for the problem at hand.
However, since  \pkg{pomp} is designed for general POMP models, methods that exploit additional model structure have yet to be implemented.
In particular, when linear, Gaussian approximations are adequate for one's purposes, or when the latent process takes values in a small, discrete set, methods that exploit these additional assumptions to advantage, such as the extended and ensemble Kalman filter methods or exact hidden-Markov-model methods, are available, but not yet as part of \pkg{pomp}.
It is the class of nonlinear, non-Gaussian POMP models with large state spaces upon which \pkg{pomp} is focused.

A POMP model may be characterized by the transition density for the Markov process and the measurement density\footnote{We use the term ``density'' in this article encompass both the continuous and discrete cases.
Thus, in the latter case, i.e., when state variables and/or measured quantities are discrete, one could replace ``probability density function'' with ``probability mass function''.}.
However, some methods require only simulation from the transition density whereas others require evaluation of this density.
Still other methods may not work with the model itself but with an approximation, such as a linearization.
Algorithms for which the dynamic model is specified only via a simulator are said to be {\it plug-and-play} \citep{breto09,he10}.
Plug-and-play methods can be employed once one has ``plugged'' a model simulator into the inference machinery.
Since many POMP models of scientific interest are relatively easy to simulate, the plug-and-play property facilitates data analysis.
Even if one candidate model has tractable transition probabilities, a scientist will frequently wish to consider alternative models for which these probabilities are intractable.
In a plug-and-play methodological environment, analysis of variations in the model can often be achieved by changing a few lines of the model simulator codes.
The price one pays for the flexibility of plug-and-play methodology is primarily additional computational effort, which can be substantial.
Nevertheless, plug-and-play methods implemented using \pkg{pomp} have proved capable for state of the art inference problems \citep[e.g.,][]{king08,bhadra11,shrestha11,shrestha13,Earn2012,roy13,blackwood13,blackwood13b,He2013a,Breto2014,Blake2014}.
The recent surge of interest in plug-and-play methodology for POMP models includes the development of
nonlinear forecasting \citep{ellner98},
iterated filtering \citep{ionides06,ionides15},
ensemble Kalman filtering \citep{Shaman2012},
approximate Bayesian computation (ABC) \citep{sisson07},
particle Markov chain Monte~Carlo (PMCMC) \citep{andrieu10},
probe matching \citep{kendall99},
and
synthetic likelihood \citep{wood10}.
Although the \pkg{pomp} package provides a general environment for methods with and without the plug-and-play property, development of the package to date has emphasized plug-and-play methods.


The \pkg{pomp} package is philosophically neutral as to the merits of Bayesian inference.
It enables a POMP model to be supplemented with prior distributions on parameters, and several Bayesian methods are implemented within the package.
Thus \pkg{pomp} is a convenient environment for those who wish to explore both Bayesian and non-Bayesian data analyses.

The remainder of this paper is organized as follows.
\Cref{sec:background} defines mathematical notation for POMP models and relates this to their representation as objects of \class{pomp} in the \pkg{pomp} package.
\Cref{sec:methods} introduces several of the statistical methods currently implemented in \pkg{pomp}.
\Cref{sec:examples} constructs and explores a simple POMP model, demonstrating the use of the available statistical methods.
\Cref{sec:EpidemicModel} illustrates the implementation of more complex POMPs, using a model of infectious disease transmission as an example.
Finally, \cref{sec:conclusion} discusses extensions and applications of \pkg{pomp}.

\section[POMP models and their representation in pomp]{POMP models and their representation in \pkg{pomp}}
\label{sec:background}

Let $\theta$ be a $p$-dimensional real-valued parameter, $\theta\in\R^p$.
For each value of $\theta$, let $\{X(t\giventh\theta),t\in T\}$ be a Markov process,
with $X(t\giventh\theta)$ taking values in $\R^q$.
The time index set $T\subset\mathbb{R}$ may be an interval or a discrete set.
Let $\{t_i\in T, i=1,\dots,N\}$, be the times at which $X(t\giventh\theta)$ is observed, and $t_{0}\in T$ be an initial time.
Assume $t_{0}\le t_{1}<t_{2}<\cdots<t_{N}$.
We write $X_i=X(t_i\giventh\theta)$ and $X_{i:j}=(X_{i},\ X_{i+1},\dots,X_{j})$.
The process $X_{0:N}$ is only observed by way of another process $Y_{1:N}=(Y_1,\dots,Y_N)$ with $Y_n$ taking values in  $\R^r$.
The observable random variables $Y_{1:N}$ are assumed to be conditionally independent given $X_{0:N}$.
The data, $y_{1:N}^{*}=(y_{1}^{*},\ \ldots,\ y_{N}^{*})$, are modeled as a realization of this observation process and are considered fixed.
We suppose that $X_{0:N}$ and $Y_{1:N}$ have a joint density $f_{X_{0:N},Y_{1:N}}(x_{0:n},\ y_{1:n}\giventh\theta)$.
The POMP structure implies that this joint density is determined by the initial density, $f_{X_{0}}(x_{0};\theta)$, together with the conditional transition probability density, $f_{X_{n}|X_{n-1}}(x_{n}\given x_{n-1}\giventh\theta)$, and the measurement density, $f_{Y_{n}|X_{n}}(y_{n}\given x_{n}\giventh\theta)$, for $1\leq n\leq N$.
In particular, we have
\begin{equation}\label{eq:joint-dens}
  f_{X_{0:N},Y_{1:N}}(x_{0:N},y_{1:N};\theta) = f_{X_0}(x_0;\theta)\,\prod_{n=1}^N\!f_{X_n | X_{n-1}}(x_n|x_{n-1};\theta)\,f_{Y_n|X_n}(y_n|x_n;\theta).
\end{equation}
Note that this formalism allows the transition density, $f_{X_{n}|X_{n-1}}$, and measurement density, $f_{Y_{n}|X_{n}}$, to depend explicitly on $n$.

\subsection{Implementation of POMP models}
\label{sec:implementation}

\pkg{pomp} is fully object-oriented:
in the package, a POMP model is represented by an S4 object \citep{Chambers1998,genolini08} of \class{pomp}.
Slots in this object encode the components of the POMP model, and can be filled or changed using the constructor function \code{pomp} and various other convenience functions.
Methods for the \class{pomp} class use these components to carry out computations on the model.
\Cref{tab:notation} gives the mathematical notation corresponding to the elementary methods that can be executed on a \class{pomp} object.

\begin{table}[t]
  \begin{center}
    \begin{tabular}{llll}
      \hline
      Method &Argument to the &Mathematical terminology \\
      & \code{pomp} constructor & \\
      \hline
      \code{rprocess} &\code{rprocess} &Simulate from $f_{X_n|X_{n-1}}( x_n \given x_{n-1}\giventh \theta)$\\
      \code{dprocess} &\code{dprocess} &Evaluate $f_{X_n|X_{n-1}}( x_n \given x_{n-1}\giventh \theta)$\\
      \code{rmeasure} &\code{rmeasure} &Simulate from $f_{Y_n|X_n}( y_n \given x_n\giventh \theta)$\\
      \code{dmeasure} &\code{dmeasure} &Evaluate $f_{Y_n|X_n}( y_n \given x_n\giventh \theta)$\\
      \code{rprior} &\code{rprior} &Simulate from the prior distribution $\pi(\theta)$\\
      \code{dprior} &\code{dprior} &Evaluate the prior density $\pi(\theta)$\\
      \code{init.state} &\code{initializer} &Simulate from $f_{X_0}( x_0 \giventh \theta)$\\
      \code{timezero} &\code{t0} &$t_0$\\
      \code{time} &\code{times} &$t_{1:N}$\\
      \code{obs} &\code{data} &$y^*_{1:N}$\\
      \code{states} & --- &$x_{0:N}$\\
      \code{coef} &\code{params} &$\theta$\\
      \hline
    \end{tabular}
  \end{center}
  \caption{
    Constituent methods for \class{pomp} objects and their translation into mathematical notation for POMP models.
    For example, the \code{rprocess} method is set using the \code{rprocess} argument to the \code{pomp} constructor function.
    \label{tab:notation}
  }
\end{table}

The \code{rprocess}, \code{dprocess}, \code{rmeasure}, and \code{dmeasure} arguments specify the transition probabilities $f_{X_n|X_{n-1}}( x_n \given x_{n-1}\giventh \theta)$ and measurement densities $f_{Y_n|X_n}(y_n\given x_n\giventh \theta)$.
Not all of these arguments must be supplied for any specific computation.
In particular, plug-and-play methodology by definition never uses \code{dprocess}.
An empty \slot{dprocess} slot in a \class{pomp} object is therefore acceptable unless a non-plug-and-play algorithm is attempted.
In the package, the data and corresponding measurement times are considered necessary parts of a \class{pomp} object whilst specific values of the parameters and latent states are not.
Applying the \code{simulate} function to an object of \class{pomp} returns another object of \class{pomp}, within which the data $y^*_{1:N}$ have been replaced by a stochastic realization of $Y_{1:N}$, the corresponding realization of $X_{0:N}$ is accessible via the \code{states} method, and the \code{params} slot has been filled with the value of $\theta$ used in the simulation.

To illustrate the specification of models in \pkg{pomp} and the use of the package's inference algorithms, we will use a simple example.
The \citet{Gompertz1825} model can be constructed via
\begin{knitrout}
\definecolor{shadecolor}{rgb}{1, 1, 1}\color{fgcolor}\begin{kframe}
\begin{verbatim}
R> library("pomp")
R> pompExample(gompertz)
\end{verbatim}
\end{kframe}
\end{knitrout}
which results in the creation of an object of \class{pomp}, named \code{gompertz}, in the workspace.
The structure of this model and its implementation in \pkg{pomp} is described below, in \cref{sec:examples}.
One can view the components of \code{gompertz} listed in \cref{tab:notation} by executing
\begin{knitrout}
\definecolor{shadecolor}{rgb}{1, 1, 1}\color{fgcolor}\begin{kframe}
\begin{verbatim}
R> obs(gompertz)
R> states(gompertz)
R> as.data.frame(gompertz)
R> plot(gompertz)
R> timezero(gompertz)
R> time(gompertz)
R> coef(gompertz)
R> init.state(gompertz)
\end{verbatim}
\end{kframe}
\end{knitrout}
Executing \code{pompExample()} lists other examples provided with the package.

\subsection{Initial conditions}

In some experimental situations, $f_{X_0}(x_0\giventh\theta)$ corresponds to a known experimental initialization, but in general the initial state of the latent process will have to be inferred.
If the transition density for the dynamic model, $f_{X_n|X_{n-1}}(x_n\given x_{n-1}\giventh \theta)$, does not depend on time and possesses a unique stationary distribution, it may be natural to set $f_{X_0}(x_0\giventh\theta)$ to be this stationary distribution.
Otherwise, and more commonly in the authors' experience, no clear scientifically motivated choice of $f_{X_0}(x_0\giventh\theta)$ exists and one can proceed by treating the value of $X_0$ as a parameter to be estimated.
In this case, $f_{X_0}(x_0\giventh\theta)$ concentrates at a point, the location of which depends on $\theta$.

\subsection{Covariates}

Scientifically, one may be interested in the role of a vector-valued covariate process $\{Z(t)\}$ in explaining the data.
Modeling and inference conditional on $\{Z(t)\}$ can be carried out within the general framework for nonhomogeneous POMP models, since the arbitrary densities $f_{X_n|X_{n-1}}$, $f_{X_0}$ and $f_{Y_n|X_n}$ can depend on the observed process $\{Z(t)\}$.
For example, it may be the case that $f_{X_n|X_{n-1}}(x_n\given x_{n-1}\giventh\theta)$ depends on $n$ only through $Z(t)$ for $t_{n-1}\le t\le t_{n}$.
The \code{covar} argument in the \pkg{pomp} constructor allows for time-varying covariates measured at times specified in the \code{tcovar} argument.
A example using covariates is given in \cref{sec:EpidemicModel}.

\section{Methodology for POMP models}
\label{sec:methods}

Data analysis typically involves identifying regions of parameter space within which a postulated model is statistically consistent with the data.
Additionally, one frequently desires to assess the relative merits of alternative models as explanations of the data.
Once the user has encoded one or more POMP models as objects of \class{pomp}, the package provides a variety of algorithms to assist with these data analysis goals.
\Cref{tab:methods} provides an overview of several inference methodologies for POMP models.
Each method may be categorized as full-information or feature-based, Bayesian or frequentist, and plug-and-play or not plug-and-play.

Approaches that work with the full likelihood function, whether in a Bayesian or frequentist context, can be called full-information methods.
Since low-dimensional sufficient statistics are not generally available for POMP models, methods which take advantage of favorable low-dimensional representations of the data typically lose some statistical efficiency.
We use the term ``feature-based'' to describe all methods not based on the full likelihood, since such methods statistically emphasize some features of the data over others.

Many Monte Carlo methods of inference can be viewed as algorithms for the exploration of high-dimensional surfaces.
This view obtains whether the surface in question is the likelihood surface or that of some other objective function.
The premise behind many recent methodological developments in Monte Carlo methods for POMP models is that generic stochastic numerical analysis tools, such as standard Markov chain Monte Carlo and Robbins-Monro type methods, are effective only on the simplest models.
For many models of scientific interest, therefore, methods that leverage the POMP structure are needed.
Though \pkg{pomp} has sufficient flexibility to encode arbitrary POMP models and methods and therefore also provides a platform for the development of novel POMP inference methodology,
\pkg{pomp}'s development to date has focused on plug-and-play methods.
However, the package developers welcome contributions and collaborations to further expand \pkg{pomp}'s functionality in non-plug-and-play directions also.
In the remainder of this section, we describe and discuss several inference methods, all currently implemented in the package.

\begin{table}[t]
  \begin{tabular}{l|p{0.35\linewidth}|p{0.35\linewidth}}
    \multicolumn{3}{l}{\bf (a) Plug-and-play \rule[-2mm]{0mm}{4mm}  }\tabularnewline
    \hline
    &Frequentist   & Bayesian  \tabularnewline
    \hline
    Full information&
    Iterated filtering (\code{mif}, \cref{sec:mif}) \raggedright
    &PMCMC (\code{pmcmc}, \cref{sec:pmcmc}) \raggedright \tabularnewline
    \hline
    Feature-based
    &Nonlinear forecasting (\code{nlf}, \cref{sec:nlf}), \raggedright
    &ABC (\code{abc}, \cref{sec:abc}) \raggedright \tabularnewline
    &synthetic likelihood (\code{probe.match}, \cref{sec:probe}) \raggedright
    & \tabularnewline
    \hline
    \multicolumn{3}{c}{}\tabularnewline
    \multicolumn{3}{l}{\bf (b) Not plug-and-play \rule[-2mm]{0mm}{4mm}} \tabularnewline
    \hline
    & Frequentist         & Bayesian  \tabularnewline
    \hline
    Full information
    & EM and Monte~Carlo~EM,  \raggedright
    & MCMC \raggedright \tabularnewline
    & Kalman filter \raggedright
    & \tabularnewline
    \hline
    Feature-based
    &Trajectory matching (\code{traj.match}),  \raggedright
    & Extended Kalman filter \tabularnewline
    &extended Kalman filter,  \raggedright
    & \tabularnewline
    &Yule-Walker equations  \raggedright
    & \tabularnewline
    \hline
  \end{tabular}
  \caption{
    Inference methods for POMP models.
    For those currently implemented in \pkg{pomp}, function name and a reference for  description are provided in parentheses.
    Standard Expectation-Maximization (EM) and Markov chain Monte~Carlo (MCMC) algorithms are not plug-and-play since they require evaluation of $f_{X_n|X_{n-1}}(x_n\given x_{n-1}\giventh\theta)$.
    The Kalman filter and extended Kalman filter are not plug-and-play since they cannot be implemented based on a model simulator.
    The Kalman filter provides the likelihood for a linear, Gaussian model.
    The extended Kalman filter employs a local linear Gaussian approximation which can be used for frequentist inference (via maximization of the resulting quasi-likelihood) or approximate Bayesian inference (by adding the parameters to the state vector).
    The Yule-Walker equations for ARMA models provide an example of a closed-form method of moments estimator.
  }
  \label{tab:methods}
\end{table}

\subsection{The likelihood function and sequential Monte Carlo}
\label{sec:pfilter}

\begin{algorithm}[ht]
  \caption{\textbf{Sequential Monte Carlo (SMC, or particle filter)}:
    \code{pfilter(\,P,\,Np{\argequals}$J$)}, using notation from \cref{tab:notation} where \code{P} is a \class{pomp} object with definitions for \code{rprocess}, \code{dmeasure}, \code{init.state}, \code{coef}, and \code{obs}.
    \label{alg:pfilter}
  }
  \KwIn{
    Simulator for $f_{X_n|X_{n-1}}(x_n\given x_{n-1}\giventh\theta)$;
    evaluator for $f_{Y_n|X_n}(y_n\given x_{n}\giventh\theta)$;
    simulator for $f_{X_0}(x_0\giventh\theta)$;
    parameter, $\theta$;
    data, $y^*_{1:N}$;
    number of particles, $J$.
  }
  \BlankLine
  Initialize filter particles:
  simulate ${X}_{0,j}^{F}\sim {f}_{{X}_{0}}\left(\mydot\giventh{\theta}\right)$ for $j$ in $\seq{1}{J}$.\;
  \For{$n$ in $\seq{1}{N}$}{
    Simulate for prediction:
    ${X}_{n,j}^{P}\sim {f}_{{X}_{n}|{X}_{n-1}}\big(\mydot|{X}_{n-1,j}^{F};{\theta}\big)$ for $j\ \text{in}\ \seq{1}{J}$. \nllabel{alg:pfilter:step1}\;
    Evaluate weights:
    $w(n,j)={f}_{{Y}_{n}|{X}_{n}}(y_{n}^{*}|{X}_{n,j}^{P}\giventh{\theta})$ for $j$ in $\seq{1}{J}$.\;
    Normalize weights:
    $\tilde{w}(n,j)= w(n,j)/\sum_{m=1}^{J}w(n,m)$.\;
    Apply \cref{alg:systematic} to select indices $k_{1:J}$ with $\prob{k_{j}=m} =\tilde{w}(n,m)$.\nllabel{alg:pfilter:systematic}\;
    Resample:
    set ${X}_{n,j}^{F}={X}_{n,k_{j}}^{P}$ for $j$ in $\seq{1}{J}$. \nllabel{alg:pfilter:step2} \;
    Compute conditional log likelihood: $\loglikMC_{n|1:n-1}=\log\big(J^{-1}\,\sum_{m=1}^{J}\!w(n,m)\big)$.\;
  }
  \KwOut{
    Log likelihood estimate, $\loglikMC(\theta)=\sum_{n=1}^N\loglikMC_{n|1:n-1}$;
    filter sample, $X^F_{n,1:J}$, for $n$ in $\seq{1}{N}$.
  }
  \KwCplx{$\bigO{J}$}
\end{algorithm}

The log likelihood for a POMP model is $\loglik(\theta)=\log{f_{Y_{1:N}}(y^*_{1:N}\giventh\theta)}$, which can be written as a sum of conditional log likelihoods,
\begin{equation}\label{eq:loglik:factorization}
\loglik(\theta)=\sum_{n=1}^N\!\loglik_{n|1:n-1}(\theta),
\end{equation}
where
\begin{equation}
\loglik_{n|1:n-1}(\theta)=\log f_{Y_n|Y_{1:n-1}}(y^*_n\given y^*_{1:n-1}\giventh\theta),
\end{equation}
and we use the convention that $y^*_{1:0}$ is an empty vector.
The structure of a POMP model implies the representation
\begin{equation}\label{eq:condLoglik}
\loglik_{n|1:n-1}(\theta)=\log\int \!
f_{Y_{n}|X_{n}}(y_{n}^{*}|x_{n}\giventh\theta) f_{X_n|Y_{1:n-1}}(x_n\given y^*_{1:n-1}\giventh\theta)\,
dx_{n}
\end{equation}
(cf.~\cref{eq:joint-dens}).
Although $\loglik(\theta)$ typically has no closed form, it can frequently be computed by Monte Carlo methods.
Sequential Monte Carlo (SMC) builds up a representation of $f_{X_n|Y_{1:n-1}}(x_n\given y^*_{1:n-1}\giventh\theta)$ that can be used to obtain an estimate, $\loglikMC_{n|1:n-1}(\theta)$, of $\loglik_{n|1:n-1}(\theta)$ and hence an approximation, $\loglikMC(\theta)$, to $\loglik(\theta)$.
SMC (a basic version of which is presented as \cref{alg:pfilter}), is also known as the particle filter, since it is conventional to describe the Monte Carlo sample, $\{X^F_{n,j},j\ \text{in}\ \seq{1}{J}\}$ as a swarm of particles representing $f_{X_n|Y_{1:n}}(x_n\given y^*_{1:n}\giventh\theta)$.
The swarm is propagated forward according to the dynamic model and then assimilated to the next data point.
Using an evolutionary analogy, the prediction step (\cref{alg:pfilter:step1}) mutates the particles in the swarm and the filtering step (\cref{alg:pfilter:step2}) corresponds to selection.
SMC is implemented in \pkg{pomp} in the \code{pfilter} function.
The basic particle filter in \cref{alg:pfilter} possesses the plug-and-play property.
Many variations and elaborations to SMC have been proposed;
these may improve numerical performance in appropriate situations \citep{cappe07} but typically lose the plug-and-play property.
\citet{arulampalam02}, \citet{Doucet2009}, and \citet{Kantas2015} have written excellent introductory tutorials on the particle filter and particle methods more generally.

Basic SMC methods fail when an observation is extremely unlikely given the model.
This leads to the situation that at most a few particles are consistent with the observation, in which case the effective sample size \citep{liu01a} of the Monte Carlo sample is small and the particle filter is said to suffer from \emph{particle depletion}.
Many elaborations of the basic SMC algorithm have been proposed to ameliorate this problem.
However, it is often preferable to remedy the situation by seeking a better model.
The plug-and-play property assists in this process by facilitating investigation of alternative models.

\begin{algorithm}[t]
  \caption{
    \textbf{Systematic resampling}:
    \Cref{alg:pfilter:systematic} of \cref{alg:pfilter}.
    \label{alg:systematic}
  }
  \KwIn{
    Weights, $\tilde{w}_{1:J}$, normalized so that $\sum_{j=1}^J \tilde{w}_j=1$.
  }
  \BlankLine
  Construct cumulative sum:
  $c_j=\sum_{m=1}^j \tilde{w}_m$, for $j$ in $1:J$.\;
  Draw a uniform initial sampling point:
  $U_1\sim\mathrm{Uniform}(0,J^{-1})$.\;
  Construct evenly spaced sampling points:
  $U_j=U_1 + (j-1)J^{-1}$, for $j\ \text{in}\ 2:J$.\;
  Initialize: set $p=1$.\;
  \For {$j\ \text{in}\ 1:J$}{
    \While {$U_j>c_p$}{
      Step to the next resampling index:
      set $p=p+1$.\;
    }
    Assign resampling index:
    set $k_j=p$.\;
  }
  \KwOut{Resampling indices, $k_{1:J}$.}
  \KwCplx{$\bigO{J}$}
\end{algorithm}

In \cref{alg:pfilter:systematic} of \cref{alg:pfilter}, systematic resampling (\cref{alg:systematic}) is used in preference to multinomial resampling.
\Cref{alg:systematic} reduces Monte~Carlo variability while resampling with the proper marginal probability.
In particular, if all the particle weights are equal then \cref{alg:systematic} has the appropriate behavior of leaving the particles unchanged.
As pointed out by \citep{douc05}, stratified resampling performs better than multinomial sampling and \cref{alg:systematic} is in practice comparable in performance to stratified resampling and somewhat faster.

\pagebreak

\subsection{Iterated filtering}
\label{sec:mif}

\begin{algorithm}[h]
  \caption{
    \textbf{Iterated filtering}:
    \texttt{mif(P, start{\argequals}$\theta_0$, Nmif{\argequals}$M$, Np{\argequals}$J$, rw.sd{\argequals}$\sigma_{1:p}$, ic.lag{\argequals}$L$, var.factor{\argequals}$C$, cooling.factor{\argequals}$a$)},
    using notation from \cref{tab:notation}
    where \code{P} is a \class{pomp} object with defined \code{rprocess}, \code{dmeasure}, \code{init.state}, and \code{obs} components.
    \label{alg:mif}
    }
  \KwIn{
    Starting parameter, $\theta_0$;
    simulator for $f_{X_0}(x_0\giventh\theta)$;
    simulator for $f_{X_n|X_{n-1}}(x_n\given x_{n-1}\giventh\theta)$;
    evaluator for $f_{Y_n|X_n}(y_n\given x_{n}\giventh\theta)$;
    data, $y^*_{1:N}$;
    labels, $I\subset\{1,\dots,p\}$, designating IVPs;
    fixed lag, $L$, for estimating IVPs;
    number of particles, $J$,
    number of iterations, $M$;
    cooling rate, $0<a<1$;
    perturbation scales, $\sigma_{1:p}$;
    initial scale multiplier, $C>0$.
  }
  \BlankLine
  \For {$m$ in $\seq{1}{M}$}{
    Initialize parameters: \nllabel{alg:mif:init:perturb}
    $[\Theta^F_{0,j}]_i \sim \normal\left([\theta_{m-1}]_i, (C a^{m-1} \sigma_i)^2\right)$ for $i$ in $\seq{1}{p}$, $j$ in $1\myto J$.\;
    Initialize states: \nllabel{alg:mif:initstates}
    simulate $X_{0,j}^F \sim f_{X_0}\big(\mydot;{\Theta^F_{0,j}}\big)$ for $j$ \textrm{in} $\seq{1}{J}$.\;
    Initialize filter mean for parameters:
    $\bar \theta_0=\theta_{m-1}$.\;
    Define 
    $[V_1]_i=(C^2+1)a^{2m-2}\sigma_i^2$.\;
    \For {$n$ \textrm{in} $1\myto N$}{
      Perturb parameters: \nllabel{alg:mif:perturb}
      $\big[\Theta^P_{n,j}\big]_i\sim\normal\left(\big[\Theta^F_{n-1,j}\big]_i,(a^{m-1} \sigma_i)^2\right)$ for $i \not\in I$, $j$ in $1\myto J$.\;
      Simulate prediction particles: \nllabel{alg:mif:sim}
      ${X}_{n,j}^{P}\sim {f}_{{X}_{n}|{X}_{n-1}}\big(\mydot|{X}_{n-1,j}^{F}\giventh{\Theta^P_{n,j}}\big)$ for $j$ in $1\myto J$.\;
      Evaluate weights: \nllabel{alg:mif:weights}
      $w(n,j)=f_{Y_{n}|X_{n}}(y_{n}^{*}|X_{n,j}^{P}\giventh\Theta^P_{n,j})$ for $j$ in $1\myto J$.\;
      Normalize weights: \nllabel{alg:mif:normalize}
      $\tilde{w}(n,j)= w(n,j)/\sum_{u=1}^{J}w(n,u)$.\;
      Apply \cref{alg:systematic} to select indices $k_{1:J}$ with $\prob{k_{u}=j} =\tilde{w}\left(n,j\right)$.\nllabel{alg:mif:syst}\;
      Resample particles: \nllabel{alg:mif:resample}
      $X_{n,j}^{F}=X_{n,k_{j}}^{P}$ and $\Theta_{n,j}^{F}=\Theta^P_{n,k_{j}}$ for $j$ in $1\myto J$.\;
      Filter mean: \nllabel{alg:mif:filtermean}
      $\big[\bar{\theta}_{n}\big]_i=\sum_{j=1}^J\tilde{w}(n,j)\big[\Theta^P_{n,j}\big]_i$ for $i \not\in I$.\;
      Prediction variance: \nllabel{alg:mif:predvar}
      $[V_{n+1}]_i=(a^{m-1}\sigma_i)^2 + \sum_{j}\tilde{w}(n,j)\big([\Theta^P_{n,j}]_i - [\bar{\theta}_{n}]_i\big)^2$ for $i \not\in I$.\;
    }
    Update non-IVPs: \nllabel{alg:mif:update}
    $[\theta_m]_i=[\theta_{m-1}]_i+ [V_1]_i \sum_{n=1}^{N} [V_{n}]_i^{-1}\left([\bar{\theta}_{n}]_{i}-[\bar{\theta}_{n-1}]_{i}\right)$ for $i \not\in I$.\;
    Update IVPs: \nllabel{alg:mif:updateivps}
    $[\theta_m]_i=\frac{1}{J}\sum_{j}\big[\Theta_{L,j}^F\big]_i$ for $i \in I$.\;
  }
  \KwOut{Monte~Carlo maximum likelihood estimate, $\theta_M$.}
  \KwCplx{$\bigO{J M}$}
\end{algorithm}

Iterated filtering techniques maximize the likelihood obtained by SMC \citep{ionides06,ionides11,ionides15}.
The key idea of iterated filtering is to replace the model we are interested in fitting---which has time-invariant parameters---with a model that is just the same except that its parameters take a random walk in time.
Over multiple repetitions of the filtering procedure, the intensity of this random walk approaches zero and the modified model approaches the original one.
Adding additional variability in this way has four positive effects:
\begin{enumerate}[{A}1.]
\item It smooths the likelihood surface, which facilitates optimization.
\item It combats particle depletion by adding diversity to the population of particles.
\item The additional variability can be exploited to explore the likelihood surface and estimate of the gradient of the (smoothed) likelihood, based on the same filtering procedure that is required to estimate of the value of the likelihood \citep{ionides06,ionides11}.
\item It preserves the plug-and-play property, inherited from the particle filter.
\end{enumerate}
Iterated filtering is implemented in the \code{mif} function.
By default, \code{mif} carries out the procedure of \citet{ionides06}.
The improved iterated filtering algorithm (IF2) of \citet{ionides15} has shown promise.
A limited version of IF2 is available via the \code{method="mif2"} option;
a full version of this algorithm will be released soon.
In all iterated filtering methods, by analogy with annealing, the random walk intensity can be called a temperature, which is decreased according to a prescribed cooling schedule.
One strives to ensure that the algorithm will freeze at the maximum of the likelihood as the temperature approaches zero.

The perturbations on the parameters in \cref{alg:mif:init:perturb,alg:mif:perturb} of \cref{alg:mif} follow a normal distribution, with each component, $[\theta]_i$, of the parameter vector perturbed independently.
Neither normality nor independence are necessary for iterated filtering, but, rather than varying the perturbation distribution, one can transform the parameters to make these simple algorithmic choices reasonable.

\Cref{alg:mif} gives special treatment to a subset of the components of the parameter vector termed initial value parameters (IVPs), which arise when unknown initial conditions are modeled as parameters.
These IVPs will typically be inconsistently estimable as the length of the time series increases, since for a stochastic process one expects only early time points to contain information about the initial state.
Searching the parameter space using time-varying parameters is inefficient in this situation, and so \cref{alg:mif} perturbs these parameters only at time zero.

\Cref{alg:mif:perturb,alg:mif:sim,alg:mif:weights,alg:mif:normalize,alg:mif:syst,alg:mif:resample} of \cref{alg:mif} are exactly an application of SMC (\cref{alg:pfilter}) to a modified POMP model in which the parameters are added to the state space.
This approach has been used in a variety of previously proposed POMP methodologies \citep{kitagawa98,janeliu01,wan00} but iterated filtering is distinguished by having theoretical justification for convergence to the maximum likelihood estimate \citep{ionides11}.

\subsection{Particle Markov chain Monte Carlo}
\label{sec:pmcmc}

\begin{algorithm}[h]
\caption{
  \textbf{Particle Markov Chain Monte Carlo}:
  \texttt{pmcmc(P, start{\argequals}$\theta_0$, Nmcmc{\argequals}$M$, Np{\argequals}$J$, proposal{\argequals}$q$)},
  using notation from \cref{tab:notation}
  where \code{P} is a \class{pomp} object with defined methods for
  \code{rprocess}, \code{dmeasure}, \code{init.state}, \code{dprior}, and \code{obs}.  The supplied \code{proposal} samples from a symmetric, but otherwise arbitrary, MCMC proposal distribution, $q(\theta^P\given\theta)$.
  \label{alg:pmcmc}
}
\KwIn{
  Starting parameter, $\theta_0$;
  simulator for $f_{X_0}(x_0\given\theta)$;
  simulator for $f_{X_n|X_{n-1}}(x_n\given x_{n-1}\giventh\theta)$;
  evaluator for $f_{Y_n|X_n}(y_n\given x_{n}\giventh\theta)$;
  simulator for $q(\theta^P\given\theta)$;
  data, $y^*_{1:N}$;
  number of particles, $J$;
  number of filtering operations, $M$;
  perturbation scales, $\sigma_{1:p}$;
  evaluator for prior, $f_{\Theta}(\theta)$.
}
\BlankLine
Initialization: compute $\loglikMC(\theta_0)$ using \cref{alg:pfilter} with $J$ particles.\;
\For {$m$ in $\seq{1}{M}$}{
  Draw a parameter proposal, $\theta^P_m$, from the proposal distribution:
  $\Theta^P_m \sim q\left(\mydot\given\theta_{m-1}\right)$.\;
  Compute $\loglikMC(\theta^P_m)$ using \cref{alg:pfilter} with $J$ particles.\;
  Generate $U\sim\mathrm{Uniform}(0,1).$\;
  Set $\big(\theta_m,\loglikMC(\theta_m)\big)=\begin{cases}
  \big(\theta^P_m,\loglikMC(\theta^P_m)\big), &\text{if } U<\displaystyle\frac{f_\Theta(\theta^P_m)\exp(\loglikMC(\theta^P_m))}{f_\Theta(\theta_{m-1})\exp(\loglikMC(\theta_{m-1}))},\\
  \big(\theta_{m-1},\loglikMC(\theta_{m-1})\big),&\text{otherwise.}
  \end{cases}$\;
}
\KwOut{
  Samples, $\theta_{1:M}$, representing the posterior distribution, $f_{\Theta|Y_{1:N}}(\theta\given y_{1:N}^*)$.
}
\KwCplx{$\bigO{J M}$}
\end{algorithm}

Full-information plug-and-play Bayesian inference for POMP models is enabled by particle Markov chain Monte Carlo (PMCMC) algorithms \citep{andrieu10}.
PMCMC methods combine likelihood evaluation via SMC with MCMC moves in the parameter space.
The simplest and most widely used PMCMC algorithm, termed particle marginal Metropolis-Hastings (PMMH), is based on the observation that the unbiased likelihood estimate provided by SMC can be plugged in to the Metropolis-Hastings update procedure to give an algorithm targeting the desired posterior distribution for the parameters \citep{andrieu09}.
PMMH is implemented in \code{pmcmc}, as described in \cref{alg:pmcmc}.
In part because it gains only a single likelihood evaluation from each particle-filtering operation, PMCMC can be computationally relatively inefficient \citep{bhadra10,ionides15}.
Nevertheless, its invention introduced the possibility of full-information plug-and-play Bayesian inferences in some situations where they had been unavailable.

\subsection{Synthetic likelihood of summary statistics}
\label{sec:probe}

\begin{algorithm}
  \caption{
    \textbf{Synthetic likelihood evaluation}:
    \protect
    \texttt{probe(P, nsim{\argequals}$J$, probes{\argequals}$\featureFunc$)},
    using notation from \cref{tab:notation}
    where \code{P} is a \class{pomp} object with defined methods for
    \code{rprocess}, \code{rmeasure}, \code{init.state}, and \code{obs}.
    \label{alg:probe}
  }
  \KwIn{
    Simulator for $f_{X_{n}|X_{n-1}}(x_{n}\given x_{n-1}\giventh\theta)$;
    simulator for $f_{X_{0}}(x_{0}\giventh\theta)$;
    simulator for $f_{Y_n|X_n}(y_n\given x_{n}\giventh\theta)$;
    parameter, $\theta$;
    data, $y^*_{1:N}$;
    number of simulations, $J$;
    vector of summary statistics or \emph{probes}, $\featureFunc=(\featureFunc_1,\dots,\featureFunc_d)$.
    }
  \BlankLine
  Compute observed probes:
  $s^*_{i}=\featureFunc_i(y^*_{1:N})$ for  $i$ in $\seq{1}{d}$.\;
  Simulate $J$ datasets:
  $Y^j_{1:N}\sim f_{Y_{1:N}}(\mydot \giventh \theta)$ for $j$ in $\seq{1}{J}$.\;
  Compute simulated probes:
  $s_{ij}=\featureFunc_i(Y^j_{1:N})$ for  $i$ in $1\myto d$ and $j$ in $\seq{1}{J}$.\;
  Compute sample mean:
  $\mu_i=J^{-1}\sum_{j=1}^Js_{ij}$ for  $i$ in $\seq{1}{d}$.\;
  Compute sample covariance:
  $\Sigma_{ik}=(J-1)^{-1}\sum_{j=1}^J (s_{ij}-\mu_i)(s_{kj}-\mu_k)$ for $i$ and $k$ in $\seq{1}{d}$.\;
  Compute the log synthetic likelihood:
  \begin{equation}
    \loglikMC_{\featureFunc}(\theta)=-\frac{1}{2}\,(s^*-\mu)^{T}{\Sigma}^{-1}(s^*-\mu)-\frac{1}{2}\,\log|\Sigma|-\frac{d}{2}\,\log(2\pi).
  \end{equation}\;
  \KwOut{Synthetic likelihood, $\synloglikMC(\theta)$.}
  \KwCplx{$\bigO{J}$}
\end{algorithm}

Some motivations to estimate parameter based on features rather than the full likelihood include
\begin{enumerate}[{B}1.]
\item \label{whyFeature2}
  Reducing the data to sensibly selected and informative low-dimensional summary statistics may have computational advantages \citep{wood10}.
\item \label{whyFeature1}
  The scientific goal may be to match some chosen characteristics of the data rather than all aspects of it.
  Acknowledging the limitations of all models, this limited aspiration may be all that can reasonably be demanded  \citep{kendall99,Wood2001a}.
\item \label{whyFeature3} In conjunction with full-information methodology, consideration of individual features has diagnostic value to determine which aspects of the data are driving the full-information inferences \citep{reuman06}.
\item  \label{whyFeature4} Feature-based methods for dynamic models typically do not require the POMP model structure.
However, that benefit is outside the scope of the \pkg{pomp} package.
\item \label{whyFeature5} Feature-based methods are typically \emph{doubly plug-and-play}, meaning that they require simulation, but not evaluation, for both the latent process transition density and the measurement model.
\end{enumerate}
When pursuing goal~B\ref{whyFeature2}, one aims to find summary statistics which are as close as possible to sufficient statistics for the unknown parameters.
Goals~B\ref{whyFeature1} and~B\ref{whyFeature3} deliberately look for features which discard information from the data;
in this context the features have been called probes \citep{kendall99}.
The features are denoted by a collection of functions, $\featureFunc=(\featureFunc_1,\dots,\featureFunc_d)$, where each $\featureFunc_i$ maps an observed time series to a real number.
We write $S=(S_1,\dots,S_d)$ for the vector-valued random variable with $S=\featureFunc(Y_{1:N})$, with $f_{S}(s\giventh\theta)$ being the corresponding joint density.
The observed feature vector is $s^*$ where $s^*_i=\featureFunc_i(y^*_{1:N})$, and for any parameter set one can look for parameter values for which typical features for simulated data match the observed features.
One can define a likelihood function, $\synloglik(\theta)=f_{S}(s^*\giventh\theta)$.
Arguing that $S$ should be approximately multivariate normal, for suitable choices of the features, \citet{wood10} proposed using simulations to construct a multivariate normal approximation to $\synloglik(\theta)$, and called this a \emph{synthetic likelihood}.

Simulation-based evaluation of a feature matching criterion is implemented by \code{probe} (\cref{alg:probe}).
The feature matching criterion requires a scale, and a natural scale to use is the empirical covariance of the simulations.
Working on this scale, as implemented by \code{probe}, there is no substantial difference between the probe approaches of \citet{kendall99} and \citet{wood10}.
Numerical optimization of the synthetic likelihood is implemented by \code{probe.match}, which offers a choice of the subplex method \citep{Rowan1990,subplex} or any method provided by \code{optim} or the \pkg{nloptr} package \citep{Johnson2014,Ypma2014}.

\subsection{Approximate Bayesian computation (ABC)}
\label{sec:abc}

\begin{algorithm}
  \caption{
    \textbf{Approximate Bayesian Computation}:
    \protect
    \texttt{abc(P, start{\argequals}$\theta_0$, Nmcmc{\argequals}$M$, probes{\argequals}$\featureFunc$, scale{\argequals}$\tau_{1:d}$, proposal{\argequals}$q$, epsilon{\argequals}$\epsilon$)},
    using notation from \cref{tab:notation}, where
    \code{P} is a \class{pomp} object with defined methods for
    \code{rprocess}, \code{rmeasure}, \code{init.state}, \code{dprior}, and \code{obs}.
    \label{alg:abc}
  }
  \KwIn{
    Starting parameter, $\theta_0$;
    simulator for $f_{X_0}(x_0\giventh\theta)$;
    simulator for $f_{X_n|X_{n-1}}(x_n\given x_{n-1}\giventh\theta)$;
    simulator for $f_{Y_n|X_n}(y_n\given x_{n}\giventh\theta)$;
    simulator for $q(\theta^P\given\theta)$;
    data, $y^*_{1:N}$;
    number of proposals, $M$;
    vector of probes, $\featureFunc=(\featureFunc_1,\dots,\featureFunc_d)$;
    perturbation scales, $\sigma_{1:p}$;
    evaluator for prior, $f_{\Theta}(\theta)$; feature scales, $\tau_{1:d}$;
    tolerance, $\epsilon$.
  }
  \BlankLine
  Compute observed probes:
  $s^*_{i}=\featureFunc_i(y^*_{1:N})$ for  $i$ in $1\myto d$.\;
  \For {$m$ in $1\myto M$}{
    Draw a parameter proposal, $\theta^P_m$, from the proposal distribution:
    $\Theta^P_m \sim q\left(\mydot\given\theta_{m-1}\right)$.\;
    Simulate dataset:
    $Y_{1:N}\sim f_{Y_{1:N}}(\mydot\giventh\theta^P_{m})$.\;
    Compute simulated probes:
    $s_i=\featureFunc_i(Y_{1:N})$ for  $i$ in $\seq{1}{d}$.\;
    Generate $U\sim\mathrm{Uniform}(0,1).$\;
    Set $\theta_m=\begin{cases}
    \theta^P_m,  &\text{if } \displaystyle \sum_{i=1}^d\left(\frac{s_i-s_i^*}{\tau_i}\right)^2<\epsilon^2 \text{ and }
    \displaystyle U<\frac{f_\Theta(\theta^P_m)}{f_\Theta(\theta_{m-1})},\\
    \theta_{m-1}, &\text{otherwise.}
    \end{cases}\nllabel{alg:abc:epsilon}$\;
  }
  \KwOut{
    Samples, ${\theta_{1:M}}$, representing the posterior distribution, $f_{\Theta|S_{1:d}}(\theta\given s_{1:d}^*)$.
  }
  \KwCplx{Nominally $\bigO{M}$, but performance will depend on the choice of $\epsilon$, $\tau_i$, and $\sigma_i$, as well as on the choice of probes $\featureFunc$.}
\end{algorithm}

ABC algorithms are Bayesian feature-matching techniques, comparable to the frequentist generalized method of moments \citep{Marin2012}.
The vector of summary statistics $\featureFunc$, the corresponding random variable $S$, and the value $s^*=\featureFunc(y^*_{1:N})$, are defined as in \cref{sec:probe}.
The goal of ABC is to approximate the posterior distribution of the unknown parameters given $S=s^*$.
ABC has typically been motivated by computational considerations, as in point~B\ref{whyFeature2} of \cref{sec:probe} \citep{sisson07,toni09,beaumont10}.
Points~B\ref{whyFeature1} and~B\ref{whyFeature3} also apply \citep{ratmann09}.

The key theoretical insight behind ABC algorithms is that an unbiased estimate of the likelihood can be substituted into a Markov chain Monte~Carlo algorithm to target the required posterior, the same result that justifies PMCMC \citep{andrieu09}.
However, ABC takes a different approach to approximating the likelihood.
The likelihood of the observed features, $\loglik_S(\theta)=f_{S}(s^*\giventh\theta)$, has an approximately unbiased estimate based on a single Monte~Carlo realization $Y_{1:N}\sim f_{Y_{1:N}}(\mydot\giventh\theta)$ given by
\begin{equation}
\label{eq:abc-lik}
\synloglikMC^{ABC}(\theta)=\begin{cases}
  \epsilon^{-d}B_d^{-1}\displaystyle\prod_{i=1}^d \tau_i,
  & \text{if } \displaystyle\sum_{i=1}^d\left(\frac{s_i-s_i^*}{\tau_i}\right)^2 < \epsilon^2, \\
0, & \text{otherwise,}
\end{cases}
\end{equation}
where $B_d$ is the volume of the $d$-dimensional unit ball and $\tau_i$ is a scaling chosen for the $i$th feature.
The likelihood approximation in \cref{eq:abc-lik} differs from the synthetic likelihood in \cref{alg:probe} in that only a single simulation is required.
As $\epsilon$ become small, the bias in \cref{eq:abc-lik} decreases but the Monte Carlo variability increases.
The ABC implementation \code{abc} (presented in \cref{alg:abc}) is a random walk Metropolis implementation of ABC-MCMC \citep[Algorithm 3 of][]{Marin2012}.
In the same style as iterated filtering and PMCMC, we assume a Gaussian random walk in parameter space;
the package supports alternative choices of proposal distribution.

\subsection{Nonlinear forecasting}
\label{sec:nlf}

\begin{algorithm}[h]
  \caption{
    \textbf{Simulated quasi log likelihood for NLF}.
    Pseudocode for the  quasi-likelihood function optimized by \texttt{nlf(\,P,\,start{\argequals}$\theta_0$, nasymp{\argequals}$J$, nconverge{\argequals}$B$, nrbf{\argequals}$K$, lags{\argequals}$\lags_{1:L}$)}.
    Using notation from \cref{tab:notation}, \code{P} is a \class{pomp} object with defined methods for \code{rprocess}, \code{rmeasure}, \code{init.state}, and \code{obs}.
    \label{alg:nlf}
  }
  \KwIn{
    Simulator for $f_{X_{n}|X_{n-1}}(x_{n}\given x_{n-1}\giventh\theta)$;
    simulator for $f_{X_{0}}(x_{0}\giventh\theta)$;
    simulator for $f_{Y_n|X_n}(y_n\given x_{n}\giventh\theta)$;
    parameter, $\theta$;
    data, $y^*_{1:N}$;
    collection of lags, $\lags_{1:L}$;
    length of discarded transient, $B$;
    length of simulation, $J$;
    number of radial basis functions, $K$.
  }
  Simulate long stationary time series:
  $Y_{1:(B+J)}\sim f_{Y_{1:(B+J)}}(\mydot\giventh\theta)$.\;
  Set $Y_{\min}=\min\{Y_{(B+1):(B+J)}\}$, $Y_{\max}=\max\{Y_{(B+1):(B+J)}\}$ and $R=Y_{\max}-Y_{\min}$.\;
  Locations for basis functions:
  $m_k=Y_{\min} + R\times [1.2\times (k-1)(K-1)^{-1}-0.1]$ for $k$ in $\seq{1}{K}$.\label{alg:nlf:rbf:params1}\;
  Scale for basis functions:
  $s=0.3\times R$ \label{alg:nlf:rbf:params2}.\;
  Define radial basis functions:
  $f_k(x)=\exp\{(x-m_k)^2/2s^2\}$ for $k$ in $\seq{1}{K}$.\label{alg:nlf:rbf}\;
  Define prediction function:
  $H(y_{n-\lags_1},y_{n-\lags_2},\dots,y_{n-\lags_L})=\sum_{j=1}^L\sum_{k=1}^K a_{jk}f_k(y_{n-\lags_j})$.\;
  Compute $\{a_{jk}: j\in\seq{1}{L}, k\in\seq{1}{K}\}$ to minimize
  \begin{equation}
    \hat\sigma^2=\frac{1}{J} \sum_{n=B+1}^{B+J} \big[ Y_n - H(Y_{n-\lags_1},Y_{n-\lags_2},\dots,Y_{n-\lags_L})\big]^2.
  \end{equation}\;
  Compute the simulated quasi log likelihood:
  \begin{equation}
    \loglikMC_{Q}(\theta)=-\frac{N-\maxlag}{2}\log2\pi\hat\sigma^2-\sum_{n=1+\maxlag}^{N}\!\frac{\big[y_n^*-H(y^*_{n-\lags_1},y^*_{n-\lags_2},\dots,y^*_{n-\lags_L})\big]^2}{2\hat\sigma^2},
  \end{equation}
  where $\maxlag=\max(\lags_{1:L}).$\;
  \KwOut{Simulated quasi log likelihood, $\loglikMC_Q(\theta)$.}
  \KwCplx{$\bigO{B}+\bigO{J}$}
\end{algorithm}


Nonlinear forecasting (NLF) uses simulations to build up an approximation to the one-step prediction distribution that is then evaluated on the data.
We saw in \cref{sec:pfilter} that SMC evaluates the prediction density for the observation, $f_{Y_n|Y_{1:{n-1}}}(y_n^*\given y_{1:n-1}^*\giventh \theta)$, by first building an approximation to the prediction density of the latent process, $f_{X_n|Y_{1:{n-1}}}(x_n\given y_{1:n-1}^*\giventh \theta)$.
NLF, by contrast, uses simulations to fit a linear regression model which predicts $Y_n$ based on a collection of $L$ lagged variables, $\{Y_{n-\lags_1},\dots,Y_{n-\lags_L}\}$.
The prediction errors when this model is applied to the data give rise to a quantity called the quasi-likelihood, which behaves for many purposes like a likelihood \citep{smith93}.
The implementation in \code{nlf} maximises the quasi-likelihood computed in \cref{alg:nlf}, using the \code{subplex} method \citep{Rowan1990,subplex} or any other optimizer offerered by \code{optim}.
The construction of the quasi-likelihood in \code{nlf} follows the specific recommendations of \citet{kendall05}.
In particular, the choice of radial basis functions, $f_k$, in \cref{alg:nlf:rbf} and the specification of $m_k$ and $s$ in \cref{alg:nlf:rbf:params1,alg:nlf:rbf:params2} were proposed by \citet{kendall05} based on trial and error.
The quasi-likelihood is mathematically most similar to a likelihood when $\min(\lags_{1:L})=1$, so that $\loglik_{Q}(\theta)$ approximates the factorization of the likelihood in \cref{eq:loglik:factorization}.
With this in mind, it is natural to set $\lags_{1:L}=1:L$.
However, \citet{kendall05} found that a two-step prediction criterion, with $\min(\lags_{1:L})=2$, led to improved numerical performance.
It is natural to ask when one might choose to use quasi-likelihood estimation in place of full likelihood estimation implemented by SMC.
Some considerations follow, closely related to the considerations for synthetic likelihood and ABC (\cref{sec:abc,sec:probe}).
\begin{enumerate}[{C}1.]
\item \label{whyNLF1}
NLF benefits from stationarity since (unlike SMC) it uses all time points in the simulation to build a prediction rule valid at all time points.
Indeed, NLF has not been considered applicable for non-stationary models and, on account of this, \code{nlf} is not appropriate if the model includes time-varying covariates.
An intermediate scenario between stationarity and full non-stationarity is seasonality, where the dynamic model is forced by cyclical covariates, and this is supported by \code{nlf} (cf.~B\ref{whyFeature2} in \cref{sec:probe}).
\item  \label{whyNLF2}  Potentially, quasi-likelihood could be preferable to full likelihood in some situations.
  It has been argued that a two-step prediction criterion may sometimes be more robust than the likelihood to model misspecification \citep{xia11} (cf.~B\ref{whyFeature1}).
\item\label{whyNLF3}
  Arguably, two-step prediction should be viewed as a diagnostic tool that can be used to complement full likelihood analysis rather than replace it \citep{ionides11-statSci} (cf.~B\ref{whyFeature3}).
\item \label{whyNLF4} NLF does not require that the model be Markovian (cf.~B\ref{whyFeature4}), although the \pkg{pomp} implementation, \code{nlf}, does.
\item \label{whyNLF5} NLF is doubly plug-and-play (cf.~B\ref{whyFeature5}).
\item\label{whyNLF6} The regression surface reconstruction carried out by NLF does not scale well with the dimension of the observed data.
NLF is recommended only for low-dimensional time series observations.
\end{enumerate}
NLF can be viewed as an estimating equation method, and so standard errors can be computed by standard sandwich estimator or bootstrap techniques \citep{kendall05}.
The optimization in NLF is typically carried out with a fixed seed for the random number generator, so the simulated quasi-likelihood is a deterministic function. If \code{rprocess} depends smoothly on the random number sequence and on the parameters, and the number of calls to the random number generator does not depend on the parameters, then fixing the seed results in a smooth objective function.
However, some common components to model simulators, such as \code{rnbinom}, make different numbers of calls to the random number generator depending on the arguments, which introduces nonsmoothness into the objective function.


\section{Model construction and data analysis: Simple examples}
\label{sec:examples}

\subsection{A first example: The Gompertz model}
\label{sec:gompertz:setup}

The plug-and-play methods in \pkg{pomp} were designed to facilitate data analysis based on complicated models, but we will first demonstrate the basics of \pkg{pomp} using simple discrete-time models, the Gompertz and Ricker models for population growth \citep{Reddingius1971,Ricker1954}.
The Ricker model will be introduced in \cref{sec:ricker:setup} and used in \cref{sec:ricker:probe.match}; the remainder of \cref{sec:examples} will use the Gompertz model.
The Gompertz model postulates that the density, $X_{t+\dlta{t}}$, of a population of organisms at time $t+\dlta{t}$ depends on the density, $X_{t}$, at time $t$ according to
\begin{equation}
  \label{eq:gompertz1}
  X_{t+\dlta{t}}=K^{1-e^{-r\,\dlta{t}}}\,X_{t}^{e^{-r\,\dlta{t}}}\,\varepsilon_{t}.
\end{equation}
In \cref{eq:gompertz1}, $K$ is the carrying capacity of the population, $r$ is a positive parameter, and the $\varepsilon_{t}$ are independent and identically-distributed lognormal random variables with $\log\varepsilon_t\sim\normal(0,\sigma^2)$.
Additionally, we will assume that the population density is observed with errors in measurement that are lognormally distributed:
\begin{equation}
  \label{eq:gompertz-obs}
  \log{Y_{t}}\;\sim\;\normal\left(\log{X_{t}},\tau^2\right).
\end{equation}
Taking a logarithmic transform of \cref{eq:gompertz1} gives
\begin{equation}
  \label{eq:gompertz2}
  \log{X_{t+\dlta{t}}}\;\sim\;\normal\left(\left(1-e^{-r\,\dlta{t}}\right)\,\log{K}+e^{-r\,\dlta{t}}\,\log{X_t},\sigma^2\right).
\end{equation}
On this transformed scale, the model is linear and Gaussian and so we can obtain exact values of the likelihood function by applying the Kalman filter.
Plug-and-play methods are not strictly needed;
this example therefore allows us to compare the results of generally applicable plug-and-play methods with exact results from the Kalman filter.
Later we will look at the Ricker model and a continuous-time model for which no such special tricks are available.

The first step in implementing this model in \pkg{pomp} is to construct an \proglang{R} object of \class{pomp} that encodes the model and the data.
This involves the specification of functions to do some or all of \code{rprocess}, \code{rmeasure}, and \code{dmeasure}, along with data and (optionally) other information.
The documentation (\code{?pomp}) spells out the usage of the \code{pomp} constructor, including detailed specifications for all its arguments and links to several examples.

To begin, we will write a function that implements the process model simulator.
This is a function that will simulate a single step ($t\to{t+\dlta{t}}$) of the unobserved process (\cref{eq:gompertz1}).
\begin{knitrout}
\definecolor{shadecolor}{rgb}{1, 1, 1}\color{fgcolor}\begin{kframe}
\begin{verbatim}
R> gompertz.proc.sim <- function(x, t, params, delta.t, ...) {
+    eps <- exp(rnorm(n = 1, mean = 0, sd = params["sigma"]))
+    S <- exp(-params["r"] * delta.t)
+    setNames(params["K"]^(1 - S) * x["X"]^S * eps, "X")
+  }
\end{verbatim}
\end{kframe}
\end{knitrout}
The translation from the mathematical description (\cref{eq:gompertz1}) to the simulator is straightforward.
When this function is called, the argument \code{x} contains the state at time \code{t}.
The parameters (including $K$, $r$, and $\sigma$) are passed in the argument \code{params}.
Notice that \code{x} and \code{params} are named numeric vectors and that the output must likewise be a named numeric vector, with names that match those of \code{x}.
The argument \code{delta.t} species the time-step size.
In this case, the time-step will be 1 unit;
we will see below how this is specified.

Next, we will implement a simulator for the observation process, \cref{eq:gompertz-obs}.
\begin{knitrout}
\definecolor{shadecolor}{rgb}{1, 1, 1}\color{fgcolor}\begin{kframe}
\begin{verbatim}
R> gompertz.meas.sim <- function(x, t, params, ...) {
+    setNames(rlnorm(n = 1, meanlog = log(x["X"]), sd = params["tau"]), "Y")
+  }
\end{verbatim}
\end{kframe}
\end{knitrout}
Again the translation from the measurement model \cref{eq:gompertz-obs} is straightforward.
When the function \code{gompertz.meas.sim} is called, the named numeric vector \code{x} will contain the unobserved states at time \code{t};
\code{params} will contain the parameters as before.
This return value will be a named numeric vector containing a single draw from the observation process (\cref{eq:gompertz-obs}).

Complementing the measurement model simulator is the corresponding measurement model density, which we implement as follows:
\begin{knitrout}
\definecolor{shadecolor}{rgb}{1, 1, 1}\color{fgcolor}\begin{kframe}
\begin{verbatim}
R> gompertz.meas.dens <- function(y, x, t, params, log, ...) {
+    dlnorm(x = y["Y"], meanlog = log(x["X"]), sdlog = params["tau"], 
+      log = log)
+  }
\end{verbatim}
\end{kframe}
\end{knitrout}
We will need this later on for inference using \code{pfilter}, \code{mif} and \code{pmcmc}.
When \code{gompertz.meas.dens} is called, \code{y} will contain the observation at time \code{t}, \code{x} and \code{params} will be as before, and the parameter \code{log} will indicate whether the likelihood (\code{log=FALSE}) or the log likelihood (\code{log=TRUE}) is required.

With the above in place, we build an object of \class{pomp} via a call to \code{pomp}:
\begin{knitrout}
\definecolor{shadecolor}{rgb}{1, 1, 1}\color{fgcolor}\begin{kframe}
\begin{verbatim}
R> gompertz <- pomp(data = data.frame(time = 1:100, Y = NA), times = "time", 
+    rprocess = discrete.time.sim(step.fun = gompertz.proc.sim, delta.t = 1), 
+    rmeasure = gompertz.meas.sim, t0 = 0)
\end{verbatim}
\end{kframe}
\end{knitrout}
The first argument (\code{data}) specifies a data frame that holds the data and the times at which the data were observed.
Since this is a toy problem, we have as yet no data;
in a moment, we will generate some simulated data.
The second argument (\code{times}) specifies which of the columns of \code{data} is the time variable.
The \code{rprocess} argument specifies that the process model simulator will be in discrete time, with each step of duration \code{delta.t} taken by the function given in the \code{step.fun} argument.
The \code{rmeasure} argument specifies the measurement model simulator function.
\code{t0} fixes $t_0$ for this model;
here we have chosen this to be one time unit prior to the first observation.

It is worth noting that implementing the \code{rprocess}, \code{rmeasure}, and \code{dmeasure} components as \proglang{R} functions, as we have done above, leads to needlessly slow computation.
As we will see below, \pkg{pomp} provides facilities for specifying the model in \proglang{C}, which can accelerate computations manyfold.

Before we can simulate from the model, we need to specify some parameter values.
The parameters must be a named numeric vector containing at least all the parameters referenced by the functions \code{gompertz.proc.sim} and \code{gompertz.meas.sim}.
The parameter vector needs to determine the initial condition $X(t_{0})$ as well.
Let us take our parameter vector to be
\begin{knitrout}
\definecolor{shadecolor}{rgb}{1, 1, 1}\color{fgcolor}\begin{kframe}
\begin{verbatim}
R> theta <- c(r = 0.1, K = 1, sigma = 0.1, tau = 0.1, X.0 = 1)
\end{verbatim}
\end{kframe}
\end{knitrout}
The parameters $r$, $K$, $\sigma$, and $\tau$ appear in \code{gompertz.proc.sim} and \code{gompertz.meas.sim}.
The initial condition $X_0$ is also given in \code{theta}.
The fact that the initial condition parameter's name ends in \code{.0} is significant:
it tells \code{pomp} that this is the initial condition of the state variable \code{X}.
This use of the \code{.0} suffix is the default behavior of \code{pomp}:
one can however parameterize the initial condition distribution arbitrarily using \code{pomp}'s optional \code{initializer} argument.

We can now simulate the model at these parameters:

\begin{knitrout}
\definecolor{shadecolor}{rgb}{1, 1, 1}\color{fgcolor}\begin{kframe}
\begin{verbatim}
R> gompertz <- simulate(gompertz, params = theta)
\end{verbatim}
\end{kframe}
\end{knitrout}
Now \code{gompertz} is identical to what it was before, except that the missing data have been replaced by simulated data.
The parameter vector (\code{theta}) at which the simulations were performed has also been saved internally to \code{gompertz}.
We can plot the simulated data via
\begin{knitrout}
\definecolor{shadecolor}{rgb}{1, 1, 1}\color{fgcolor}\begin{kframe}
\begin{verbatim}
R> plot(gompertz, variables = "Y")
\end{verbatim}
\end{kframe}
\end{knitrout}
\Cref{fig:gompertz-first-simulation-plot} shows the results of this operation.

\begin{figure}
\begin{knitrout}
\definecolor{shadecolor}{rgb}{1, 1, 1}\color{fgcolor}

\includegraphics[width=\maxwidth]{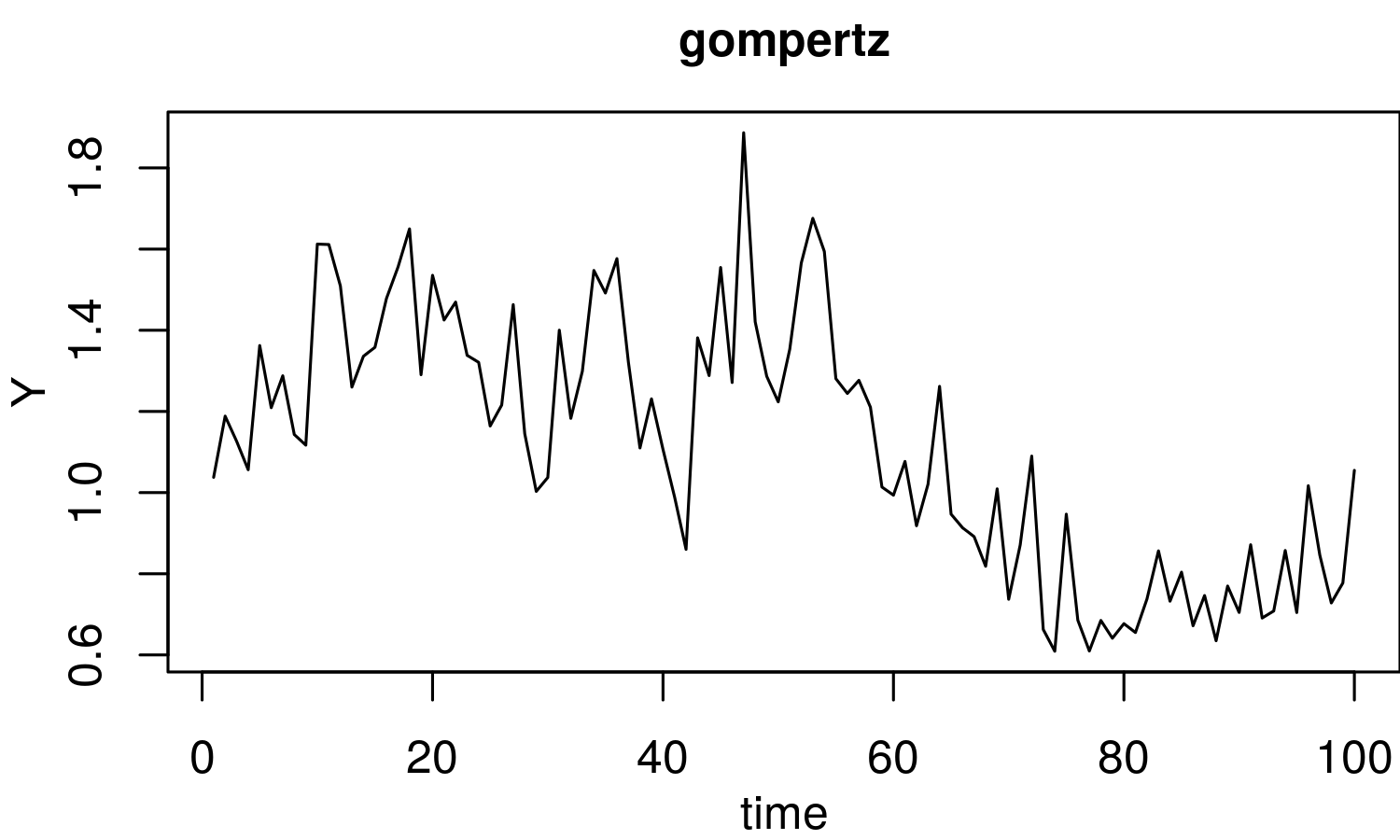} \hfill{}

\end{knitrout}
\caption{
  Simulated data from the Gompertz model (\cref{eq:gompertz1,eq:gompertz-obs}).
  This figure shows the result of executing \code{plot(gompertz, variables = "Y")}.
}
\label{fig:gompertz-first-simulation-plot}
\end{figure}

\subsection{Computing likelihood using SMC}

As discussed in \cref{sec:methods}, some parameter estimation algorithms in the \pkg{pomp} package are doubly plug-and-play in that they require only \code{rprocess} and \code{rmeasure}.
These include the nonlinear forecasting algorithm \code{nlf}, the probe-matching algorithm \code{probe.match}, and approximate Bayesian computation via \code{abc}.
The plug-and-play full-information methods in \pkg{pomp}, however, require \code{dmeasure}, i.e., the ability to evaluate the likelihood of the data given the unobserved state.
The \code{gompertz.meas.dens} above does this, but we must fold it into the \class{pomp} object in order to use it.
We can do this with another call to \code{pomp}:
\begin{knitrout}
\definecolor{shadecolor}{rgb}{1, 1, 1}\color{fgcolor}\begin{kframe}
\begin{verbatim}
R> gompertz <- pomp(gompertz, dmeasure = gompertz.meas.dens)
\end{verbatim}
\end{kframe}
\end{knitrout}
The result of the above is a new \class{pomp} object \code{gompertz} in every way identical to the one we had before, but with the measurement-model density function \code{dmeasure} now specified.

To estimate the likelihood of the data, we can use the function \code{pfilter}, an implementation of \cref{alg:pfilter}.
We must decide how many concurrent realizations (\emph{particles}) to use:
the larger the number of particles, the smaller the Monte Carlo error but the greater the computational burden.
Here, we run \code{pfilter} with 1000 particles to estimate the likelihood at the true parameters:

\begin{knitrout}
\definecolor{shadecolor}{rgb}{1, 1, 1}\color{fgcolor}\begin{kframe}
\begin{verbatim}
R> pf <- pfilter(gompertz, params = theta, Np = 1000)
R> loglik.truth <- logLik(pf)
R> loglik.truth
[1] 36.27102
\end{verbatim}
\end{kframe}
\end{knitrout}

Since the true parameters (i.e., the parameters that generated the data) are stored within the \class{pomp} object \code{gompertz} and can be extracted by the \code{coef} function, we could have done
\begin{knitrout}
\definecolor{shadecolor}{rgb}{1, 1, 1}\color{fgcolor}\begin{kframe}
\begin{verbatim}
R> pf <- pfilter(gompertz, params = coef(gompertz), Np = 1000)
\end{verbatim}
\end{kframe}
\end{knitrout}
or simply
\begin{knitrout}
\definecolor{shadecolor}{rgb}{1, 1, 1}\color{fgcolor}\begin{kframe}
\begin{verbatim}
R> pf <- pfilter(gompertz, Np = 1000)
\end{verbatim}
\end{kframe}
\end{knitrout}
Now let us compute the log likelihood at a different point in parameter space, one for which $r$, $K$, and $\sigma$ are each 50\% higher than their true values.

\begin{knitrout}
\definecolor{shadecolor}{rgb}{1, 1, 1}\color{fgcolor}\begin{kframe}
\begin{verbatim}
R> theta.guess <- theta.true <- coef(gompertz)
R> theta.guess[c("r", "K", "sigma")] <- 1.5 * theta.true[c("r", "K", "sigma")]
R> pf <- pfilter(gompertz, params = theta.guess, Np = 1000)
R> loglik.guess <- logLik(pf)
R> loglik.guess
[1] 25.19585
\end{verbatim}
\end{kframe}
\end{knitrout}
In this case, the Kalman filter computes the exact log likelihood at the true parameters to be 36.01,
while the particle filter with 1000 particles gives 36.27.
Since the particle filter gives an unbiased estimate of the likelihood, the difference is due to Monte Carlo error in the particle filter.
One can reduce this error by using a larger number of particles and/or by re-running \code{pfilter} multiple times and averaging the resulting estimated likelihoods.
The latter approach has the advantage of allowing one to estimate the Monte Carlo error itself;
we will demonstrate this in \cref{sec:gompertz:mif}.

\subsection{Maximum likelihood estimation via iterated filtering}
\label{sec:gompertz:mif}

Let us use the iterated filtering approach described in \cref{sec:mif} to obtain an approximate maximum likelihood estimate for the data in \code{gompertz}.
Since the parameters of \cref{eq:gompertz1,eq:gompertz-obs} are constrained to be positive, when estimating, we transform them to a scale on which they are unconstrained.
The following encodes such a transformation and its inverse:
\begin{knitrout}
\definecolor{shadecolor}{rgb}{1, 1, 1}\color{fgcolor}\begin{kframe}
\begin{verbatim}
R> gompertz.log.tf <- function(params, ...) log(params)
R> gompertz.exp.tf <- function(params, ...) exp(params)
\end{verbatim}
\end{kframe}
\end{knitrout}
We add these to the existing \class{pomp} object via:
\begin{knitrout}
\definecolor{shadecolor}{rgb}{1, 1, 1}\color{fgcolor}\begin{kframe}
\begin{verbatim}
R> gompertz <- pomp(gompertz, toEstimationScale = gompertz.log.tf,
+                   fromEstimationScale = gompertz.exp.tf)
\end{verbatim}
\end{kframe}
\end{knitrout}

The following codes initialize the iterated filtering algorithm at several starting points around \code{theta.true} and estimate the parameters $r$, $\tau$, and $\sigma$.

\begin{knitrout}
\definecolor{shadecolor}{rgb}{1, 1, 1}\color{fgcolor}\begin{kframe}
\begin{verbatim}
R> estpars <- c("r", "sigma", "tau")
R> library("foreach")
R> mif1 <- foreach(i = 1:10, .combine = c) %dopar% {
+    theta.guess <- theta.true
+    rlnorm(n = length(estpars), meanlog = log(theta.guess[estpars]),
+           sdlog = 1) -> theta.guess[estpars]
+    mif(gompertz, Nmif = 100, start = theta.guess, transform = TRUE,
+        Np = 2000, var.factor = 2, cooling.fraction = 0.7,
+        rw.sd = c(r = 0.02, sigma = 0.02, tau = 0.02))
+  }
R> pf1 <- foreach(mf = mif1, .combine = c) %dopar% {
+    pf <- replicate(n = 10, logLik(pfilter(mf, Np = 10000)))
+    logmeanexp(pf)
+  }
\end{verbatim}
\end{kframe}
\end{knitrout}
Note that we have set \code{transform = TRUE} in the call to \code{mif} above:
this causes the parameter transformations we have specified to be applied to enforce the positivity of parameters.
Note also that we have used the \pkg{foreach} package \citep{foreach} to parallelize the computations.

Each of the 10 \code{mif} runs ends up at a different point estimate (\cref{fig:mif-plot}).
We focus on that with the highest estimated likelihood, having evaluated the likelihood several times to reduce the Monte Carlo error in the likelihood evaluation.
The particle filter produces an unbiased estimate of the likelihood;
therefore, we will average the likelihoods, not the log likelihoods.
\begin{knitrout}
\definecolor{shadecolor}{rgb}{1, 1, 1}\color{fgcolor}\begin{kframe}
\begin{verbatim}
R> mf1 <- mif1[[which.max(pf1)]]
R> theta.mif <- coef(mf1)
R> loglik.mif <- replicate(n = 10, logLik(pfilter(mf1, Np = 10000)))
R> loglik.mif <- logmeanexp(loglik.mif, se = TRUE)
R> theta.true <- coef(gompertz)
R> loglik.true <- replicate(n = 10, logLik(pfilter(gompertz, Np = 20000)))
R> loglik.true <- logmeanexp(loglik.true, se = TRUE)
\end{verbatim}
\end{kframe}
\end{knitrout}

\begin{figure}
\begin{knitrout}
\definecolor{shadecolor}{rgb}{1, 1, 1}\color{fgcolor}

\includegraphics[width=\maxwidth]{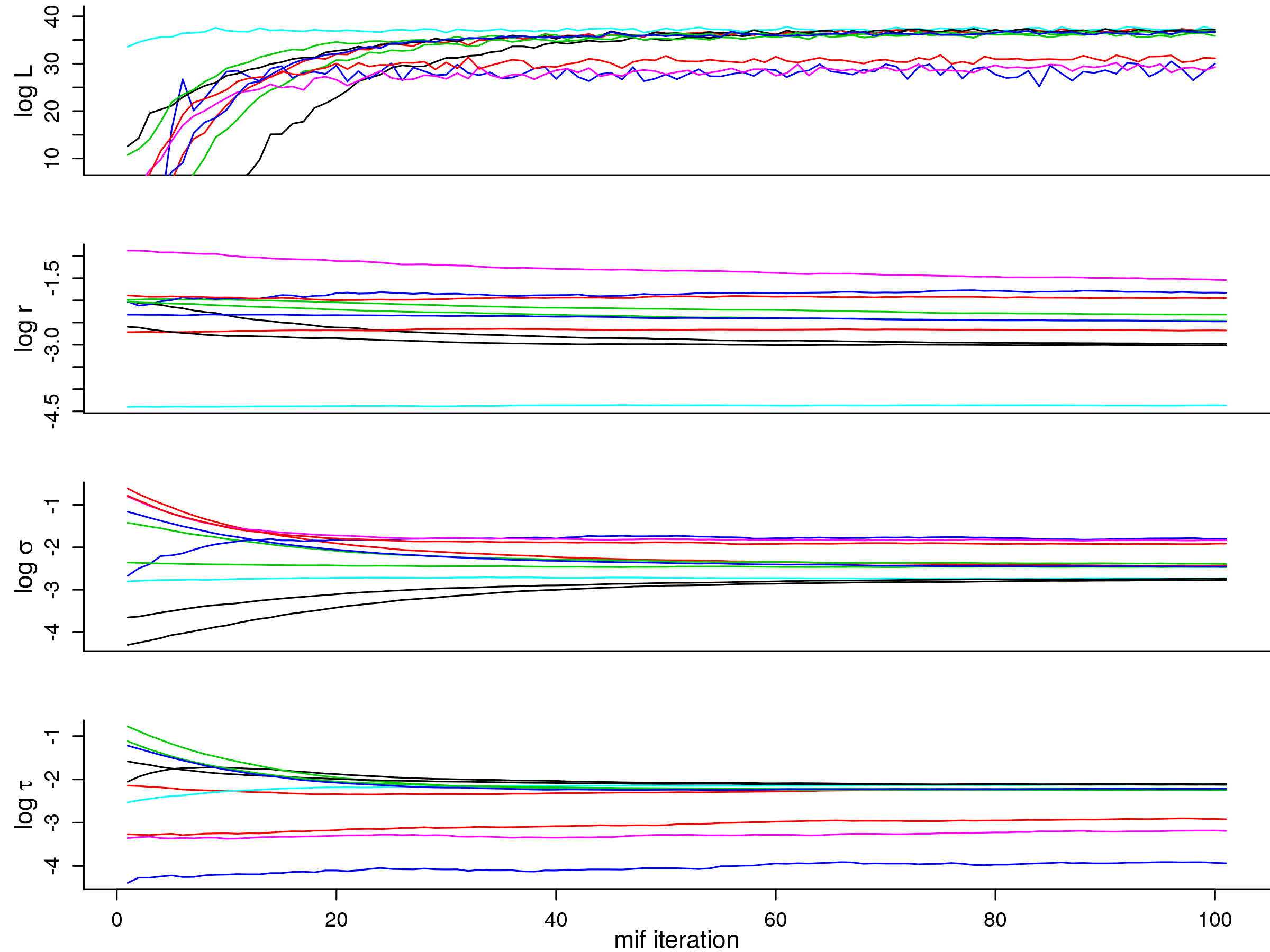} \hfill{}

\end{knitrout}
\caption{
  Convergence plots can be used to help diagnose convergence of the iterated filtering (IF) algorithm.
  These and additional diagnostic plots are produced when \code{plot} is applied to a \class{mif} or \class{mifList} object.
  \label{fig:mif-plot}
}
\end{figure}

For the calculation above, we have replicated the iterated filtering search, made a careful estimation of the log likelihood, $\loglikMC$, and its standard error using \code{pfilter} at each of the resulting point estimates, and then chosen the parameter corresponding to the highest likelihood as our numerical approximation to the MLE.
Taking advantage of the Gompertz model's tractability, we also use the Kalman filter to maximize the exact likelihood, $\loglik$, and evaluate it at the estimated MLE obtained by \code{mif}.
The resulting estimates are shown in \cref{tab:gompertz-multi-mif-table}.
Usually, the last row and column of \cref{tab:gompertz-multi-mif-table} would not be available even for a simulation study validating the inference methodology for a known POMP model.
In this case, we see that the \code{mif} procedure is successfully maximizing the likelihood up to an error of about 0.1 log units.

\begin{table}
  \begin{center}
\begin{tabular}{r|cccccc}
  \hline
 & $r$ & $\sigma$ & $\tau$ & $\loglikMC$ & s.e. & $\loglik$ \\ 
  \hline
truth & 0.1000 & 0.1000 & 0.1000 & 35.99 & 0.03 & 36.01 \\ 
  \code{mif} MLE & 0.0127 & 0.0655 & 0.1200 & 37.68 & 0.04 & 37.62 \\ 
  exact MLE & 0.0322 & 0.0694 & 0.1170 & 37.87 & 0.02 & 37.88 \\ 
   \hline
\end{tabular}

  \end{center}
  \caption{
    Results of estimating parameters $r$, $\sigma$, and $\tau$ of the Gompertz model (\cref{eq:gompertz1,eq:gompertz-obs}) by maximum likelihood using iterated filtering (\cref{alg:mif}), compared with the exact MLE and with the true value of the parameter.
    The first three columns show the estimated values of the three parameters.
    The next two columns show the log likelihood, $\loglikMC$, estimated by SMC (\cref{alg:pfilter}) and its standard error, respectively.
    The exact log likelihood, $\ell$, is shown in the rightmost column.
    An ideal likelihood-ratio $95\%$ confidence set, not usually computationally available, includes all parameters having likelihood within \code{qchisq(0.95,df=3)/2 = 3.91} of the exact MLE.
    We see that both the \code{mif} MLE and the truth are in this set.
    In this example, the \code{mif} MLE is close to the exact MLE, so it is reasonable to expect that profile likelihood confidence intervals and likelihood ratio tests constructed using the \code{mif} MLE have statistical properties similar to those based on the exact MLE.
    \label{tab:gompertz-multi-mif-table}
  }
\end{table}

\subsection{Full-information Bayesian inference via PMCMC}
\label{sec:gompertz:pmcmc}

To carry out Bayesian inference we need to specify a prior distribution on unknown parameters.
The \code{pomp} constructor function provides the \code{rprior} and \code{dprior} arguments, which can be filled with functions that simulate from and evaluate the prior density, respectively.
Methods based on random-walk Metropolis-Hastings require evaluation of the prior density (\code{dprior}), but not simulation (\code{rprior}), so we specify \code{dprior} for the Gompertz model as follows.
\begin{knitrout}
\definecolor{shadecolor}{rgb}{1, 1, 1}\color{fgcolor}\begin{kframe}
\begin{verbatim}
R> hyperparams <- list(min = coef(gompertz)/10, max = coef(gompertz) * 10)
\end{verbatim}
\end{kframe}
\end{knitrout}
\begin{knitrout}
\definecolor{shadecolor}{rgb}{1, 1, 1}\color{fgcolor}\begin{kframe}
\begin{verbatim}
R> gompertz.dprior <- function (params, ..., log) {
+    f <- sum(dunif(params, min = hyperparams$min, max = hyperparams$max,
+                   log = TRUE))
+    if (log) f else exp(f)
+  }
\end{verbatim}
\end{kframe}
\end{knitrout}
The PMCMC algorithm described in \cref{sec:pmcmc} can then be applied to draw a sample from the posterior.
Recall that, for each parameter proposal, PMCMC pays the full price of a particle-filtering operation in order to obtain the Metropolis-Hastings acceptance probability.
For the same price, iterated filtering obtains, in addition, an estimate of the derivative and a probable improvement of the parameters.
For this reason, PMCMC is relatively inefficient at traversing parameter space.
When Bayesian inference is the goal, it is therefore advisable to first locate a neighborhood of the MLE using, for example, iterated filtering.
PMCMC can then be initialized in this neighborhood to sample from the posterior distribution.
The following adopts this approach, running 5 independent PMCMC chains using a multivariate normal random-walk proposal (with diagonal variance-covariance matrix, see \code{?mvn.diag.rw}).

\begin{knitrout}
\definecolor{shadecolor}{rgb}{1, 1, 1}\color{fgcolor}\begin{kframe}
\begin{verbatim}
R> pmcmc1 <- foreach(i=1:5,.combine=c) %dopar% {
+    pmcmc(pomp(gompertz, dprior = gompertz.dprior), start = theta.mif,
+          Nmcmc = 40000, Np = 100, max.fail = Inf,
+          proposal=mvn.diag.rw(c(r = 0.01, sigma = 0.01, tau = 0.01)))
+  }
\end{verbatim}
\end{kframe}
\end{knitrout}

\begin{figure}
\begin{knitrout}
\definecolor{shadecolor}{rgb}{1, 1, 1}\color{fgcolor}

\includegraphics[width=\maxwidth]{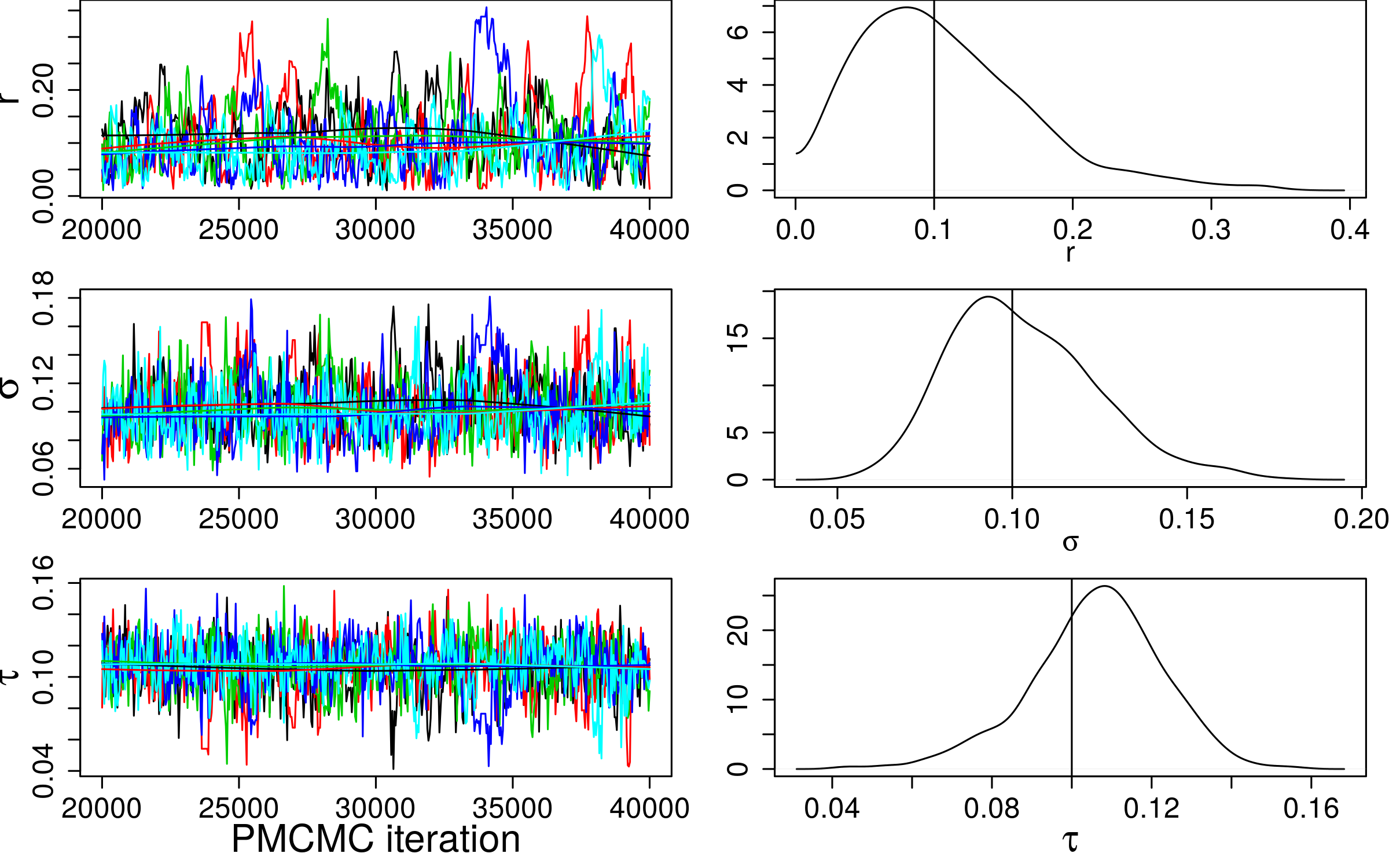} \hfill{}

\end{knitrout}
\caption{
  Diagnostic plots for the PMCMC algorithm.
  The trace plots in the left column show the evolution of 5 independent MCMC chains after a burn-in period of length 20000.
  Kernel density estimates of the marginal posterior distributions are shown at right.
  The effective sample size of the 5 MCMC chains combined is lowest for the $r$ variable and is 250:
  the use of 40000 proposal steps in this case is a modest number.
  The density plots at right show the estimated marginal posterior distributions.
  The vertical line corresponds to the true value of each parameter.
  \label{fig:pmcmc-plot}
}
\end{figure}

Comparison with the analysis of \cref{sec:gompertz:mif} reinforces the observation of \citet{bhadra10} that PMCMC can require orders of magnitude more computation than iterated filtering.
Iterated filtering may have to be repeated multiple times while computing profile likelihood plots, whereas one successful run of PMCMC is sufficient to obtain all required posterior inferences.
However, in practice, multiple runs from a range of starting points is always good practice since convergence cannot be reliably assessed on the basis of a single chain.
To verify the convergence of the approach or to compare the performance with other approaches, we can use diagnostic plots produced by the \code{plot} command (see \cref{fig:pmcmc-plot}).

\subsection{A second example: The Ricker model}
\label{sec:ricker:setup}

In \cref{sec:ricker:probe.match}, we will illustrate probe matching (\cref{sec:probe}) using a stochastic version of the Ricker map \citep{Ricker1954}.
We switch models to allow direct comparison with \citet{wood10}, whose synthetic likelihood computations are reproduced below.
In particular, the results of \cref{sec:ricker:probe.match} demonstrate frequentist inference using synthetic likelihood and also show that the full likelihood is both numerically tractable and reasonably well behaved, contrary to the claim of \citet{wood10}.
We will also take the opportunity to demonstrate features of \pkg{pomp} that allow acceleration of model codes through the use of \proglang{R}'s facilities for compiling and dynamically linking \proglang{C} code.

The Ricker model is another discrete-time model for the size of a population.
The population size, $N_t$, at time $t$ is postulated to obey
\begin{equation}\label{eq:ricker-process}
  N_{t+1}=r\,N_t\,\exp(-N_t+e_t),\qquad e_t\!\sim\!\normal\left(0,\sigma^2\right).
\end{equation}
In addition, we assume that measurements, $Y_t$, of $N_t$ are themselves noisy, according to
\begin{equation}\label{eq:ricker-measure}
  Y_t\!\sim\!\mathrm{Poisson}(\phi\,N_t),
\end{equation}
where $\phi$ is a scaling parameter.
As before, we will need to implement the model's state-process simulator (\code{rprocess}).
We have the option of writing these functions in \proglang{R}, as we did with the Gompertz model.
However, we can realize manyfold speed-ups by writing these in \proglang{C}.
In particular, \pkg{pomp} allows us to write snippets of \proglang{C} code that it assembles, compiles, and dynamically links into a running \proglang{R} session.
To begin the process, we will write snippets for the \code{rprocess}, \code{rmeasure}, and \code{dmeasure} components.
\begin{knitrout}
\definecolor{shadecolor}{rgb}{1, 1, 1}\color{fgcolor}\begin{kframe}
\begin{verbatim}
R> ricker.sim <- "
+     e = rnorm(0, sigma);
+     N = r * N * exp(-N + e);
+  "
R> ricker.rmeas <- "
+     y = rpois(phi * N);
+  "
R> ricker.dmeas <- "
+     lik = dpois(y, phi * N, give_log);
+  "
\end{verbatim}
\end{kframe}
\end{knitrout}
Note that, in this implementation, both $N$ and $e$ are state variables.
The logical flag \code{give_log} requests the likelihood when FALSE, the log likelihood when TRUE.
Notice that, in these snippets, we never declare the variables;
\pkg{pomp} will construct the appropriate declarations automatically.

In a similar fashion, we can add transformations of the parameters to enforce constraints.
\begin{knitrout}
\definecolor{shadecolor}{rgb}{1, 1, 1}\color{fgcolor}\begin{kframe}
\begin{verbatim}
R> log.trans <- "
+     Tr = log(r);
+     Tsigma = log(sigma);
+     Tphi = log(phi);
+     TN_0 = log(N_0);"
R> exp.trans <- "
+     Tr = exp(r);
+     Tsigma = exp(sigma);
+     Tphi = exp(phi);
+     TN_0 = exp(N_0);"
\end{verbatim}
\end{kframe}
\end{knitrout}
Note that in the foregoing \proglang{C} snippets, the prefix \code{T} designates the transformed version of the parameter.
A full set of rules for using \code{Csnippet}s, including illustrative examples, is given in the package help system (\code{?Csnippet}).

Now we can construct a \class{pomp} object as before and fill it with simulated data:
\begin{knitrout}
\definecolor{shadecolor}{rgb}{1, 1, 1}\color{fgcolor}\begin{kframe}
\begin{verbatim}
R> pomp(data = data.frame(time = seq(0, 50, by = 1), y = NA),
+       rprocess = discrete.time.sim(step.fun = Csnippet(ricker.sim),
+         delta.t = 1), rmeasure = Csnippet(ricker.rmeas),
+       dmeasure = Csnippet(ricker.dmeas),
+       toEstimationScale = Csnippet(log.trans),
+       fromEstimationScale = Csnippet(exp.trans),
+       paramnames = c("r", "sigma", "phi", "N.0", "e.0"),
+       statenames = c("N", "e"), times = "time", t0 = 0,
+       params = c(r = exp(3.8), sigma = 0.3, phi = 10,
+         N.0 = 7, e.0 = 0)) -> ricker
R> ricker <- simulate(ricker,seed=73691676L)
\end{verbatim}
\end{kframe}
\end{knitrout}

\subsection{Feature-based synthetic likelihood maximization}
\label{sec:ricker:probe.match}

In \pkg{pomp}, probes are simply functions that can be applied to an array of real or simulated data to yield a scalar or vector quantity.
Several functions that create useful probes are included with the package, including those recommended by \citet{wood10}.
In this illustration, we will make use of these probes:
\code{probe.marginal}, \code{probe.acf}, and \code{probe.nlar}.
\code{probe.marginal} regresses the data against a sample from a reference distribution;
the probe's values are those of the regression coefficients.
\code{probe.acf} computes the auto-correlation or auto-covariance of the data at specified lags.
\code{probe.nlar} fits a simple nonlinear (polynomial) autoregressive model to the data;
again, the coefficients of the fitted model are the probe's values.
We construct a list of probes:
\begin{knitrout}
\definecolor{shadecolor}{rgb}{1, 1, 1}\color{fgcolor}\begin{kframe}
\begin{verbatim}
R> plist <- list(probe.marginal("y", ref = obs(ricker), transform = sqrt),
+                probe.acf("y", lags = c(0, 1, 2, 3, 4), transform = sqrt),
+                probe.nlar("y", lags = c(1, 1, 1, 2), powers = c(1, 2, 3, 1),
+                           transform = sqrt))
\end{verbatim}
\end{kframe}
\end{knitrout}
Each element of \code{plist} is a function of a single argument.
Each of these functions can be applied to the data in \code{ricker} and to simulated data sets.
Calling \pkg{pomp}'s function \code{probe} results in the application of these functions to the data, and to each of some large number, \code{nsim}, of simulated data sets, and finally to a comparison of the two.
[Note that probe functions may be vector-valued, so a single probe taking values in $\R^k$ formally corresponds to a collection of $k$ probe functions in the terminology of \cref{sec:probe}.]
Here, we will apply \code{probe} to the Ricker model at the true parameters and at a wild guess.
\begin{knitrout}
\definecolor{shadecolor}{rgb}{1, 1, 1}\color{fgcolor}\begin{kframe}
\begin{verbatim}
R> pb.truth <- probe(ricker, probes = plist, nsim = 1000, seed = 1066L)
R> guess <- c(r = 20, sigma = 1, phi = 20, N.0 = 7, e.0 = 0)
R> pb.guess <- probe(ricker, params = guess, probes = plist, nsim = 1000, 
+    seed = 1066L)
\end{verbatim}
\end{kframe}
\end{knitrout}
Results summaries and diagnostic plots showing the model-data agreement and correlations among the probes can be obtained by
\begin{knitrout}
\definecolor{shadecolor}{rgb}{1, 1, 1}\color{fgcolor}\begin{kframe}
\begin{verbatim}
R> summary(pb.truth)
R> summary(pb.guess)
R> plot(pb.truth)
R> plot(pb.guess)
\end{verbatim}
\end{kframe}
\end{knitrout}

\begin{figure}
\begin{knitrout}
\definecolor{shadecolor}{rgb}{1, 1, 1}\color{fgcolor}

\includegraphics[width=\textwidth]{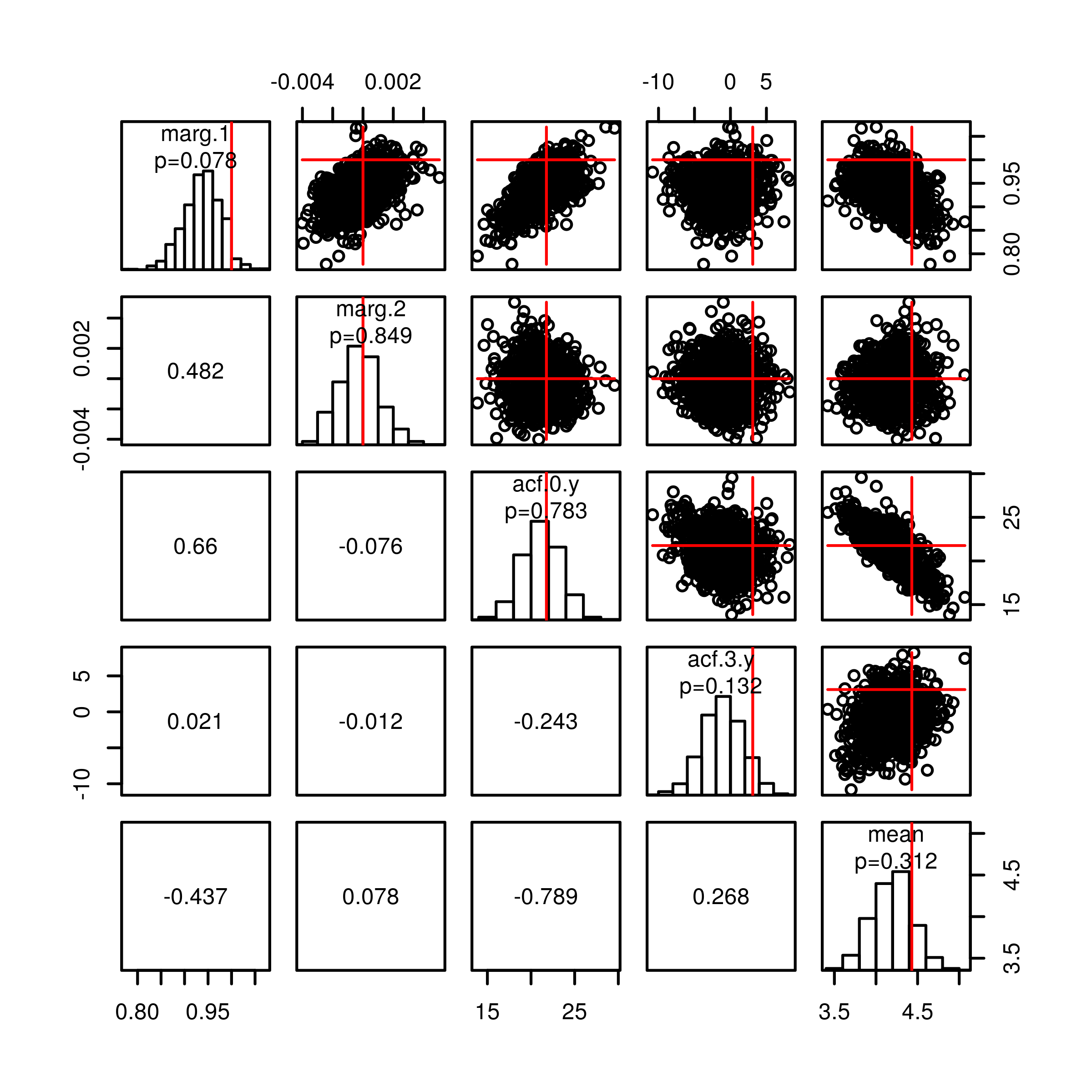} \hfill{}

\end{knitrout}
\caption{
  Results of \code{plot} on a \class{probed.pomp}-class object.
  Above the diagonal, the pairwise scatterplots show the values of the probes on each of 1000 data sets.
  The red lines show the values of each of the probes on the data.
  The panels along the diagonal show the distributions of the probes on the simulated data, together with their values on the data and a two-sided $p$~value.
  The numbers below the diagonal are the Pearson correlations between the corresponding pairs of probes.
}
\label{fig:ricker-probe-plot}
\end{figure}

An example of a diagnostic plot (using a smaller set of probes) is shown in \cref{fig:ricker-probe-plot}.
Among the quantities returned by \code{summary} is the synthetic likelihood (\cref{alg:probe}).
One can attempt to identify parameters that maximize this quantity;
this procedure is referred to in \pkg{pomp} as ``probe matching''.
Let us now attempt to fit the Ricker model to the data using probe-matching.
\begin{knitrout}
\definecolor{shadecolor}{rgb}{1, 1, 1}\color{fgcolor}\begin{kframe}
\begin{verbatim}
R> pm <- probe.match(pb.guess, est = c("r", "sigma", "phi"), transform = TRUE, 
+    method = "Nelder-Mead", maxit = 2000, seed = 1066L, reltol = 1e-08)
\end{verbatim}
\end{kframe}
\end{knitrout}

This code runs \code{optim}'s Nelder-Mead optimizer from the starting parameters \code{guess} in an attempt to maximize the synthetic likelihood based on the probes in \code{plist}.
Both the starting parameters and the list of probes are stored internally in \code{pb.guess}, which is why we need not specify them explicitly here.
While \code{probe.match} provides substantial flexibility in choice of optimization algorithm, for situations requiring greater flexibility, \pkg{pomp} provides the function \code{probe.match.objfun}, which constructs an objective function suitable for use with arbitrary optimization routines.

By way of putting the synthetic likelihood in context, let us compare the results of estimating the Ricker model parameters using probe-matching and using iterated filtering (IF), which is based on likelihood.
The following code runs 600 IF iterations starting at \code{guess}:
\begin{knitrout}
\definecolor{shadecolor}{rgb}{1, 1, 1}\color{fgcolor}\begin{kframe}
\begin{verbatim}
R> mf <- mif(ricker, start = guess, Nmif = 100, Np = 1000, transform = TRUE,
+            cooling.fraction = 0.95^50, var.factor = 2, ic.lag = 3,
+            rw.sd=c(r = 0.1, sigma = 0.1, phi = 0.1), max.fail = 50)
R> mf <- continue(mf, Nmif = 500, max.fail = 20)
\end{verbatim}
\end{kframe}
\end{knitrout}

\Cref{tab:ricker-comparison} compares parameters, Monte Carlo likelihoods ($\loglikMC$), and synthetic likelihoods ($\synloglikMC$, based on the probes in \code{plist}) at each of
\begin{inparaenum}[(a)]
\item the guess,
\item the truth,
\item the MLE from \code{mif}, and
\item the maximum synthetic likelihood estimate (MSLE) from \code{probe.match}.
\end{inparaenum}
These results demonstrate that it is possible, and indeed not difficult, to maximize the likelihood for this model, contrary to the claim of \citet{wood10}.
Since synthetic likelihood discards some of the information in the data, it is not surprising that \cref{tab:ricker-comparison} also shows the statistical inefficiency of maximum synthetic likelihood relative to that of likelihood.

\begin{table}
  \begin{center}
\begin{tabular}{r|ccccccc}
  \hline
 & $r$ & $\sigma$ & $\phi$ & $\loglikMC$ & s.e.($\loglikMC$) & $\synloglikMC$ & s.e.($\synloglikMC$) \\ 
  \hline
guess & 20.0 & 1.000 & 20.0 & -230.8 & 0.08 & -5.6 & 0.16 \\ 
  truth & 44.7 & 0.300 & 10.0 & -139.0 & 0.03 & 17.7 & 0.03 \\ 
  MLE & 45.0 & 0.186 & 10.2 & -137.2 & 0.04 & 18.0 & 0.04 \\ 
  MSLE & 42.1 & 0.337 & 11.3 & -145.7 & 0.03 & 19.4 & 0.02 \\ 
   \hline
\end{tabular}

  \end{center}
  \caption{\label{tab:ricker-comparison}
    Parameter estimation by means of maximum synthetic likelihood (\cref{alg:probe}) vs.\ by means of maximum likelihood via iterated filtering (\cref{alg:mif}).
    The row labeled ``guess'' contains the point at which both algorithms were initialized.
    That labeled ``truth'' contains the true parameter value, i.e., that at which the data were generated.
    The rows labeled ``MLE'' and ``MSLE'' show the estimates obtained using iterated filtering and maximum synthetic likelihood, respectively.
    Parameters $r$, $\sigma$, and $\tau$ were estimated; all others were held at their true values.
    The columns labeled $\loglikMC$ and $\synloglikMC$ are the Monte Carlo estimates of the log likelihood and the log synthetic likelihood, respectively;
    their Monte Carlo standard errors are also shown.
    While likelihood maximization results in an estimate for which both $\loglikMC$ and $\synloglikMC$ exceed their values at the truth,
    the value of $\loglikMC$ at the MSLE is smaller than at the truth, an indication of the relative statistical inefficiency of maximum synthetic likelihood.
  }
\end{table}

\subsection{Bayesian feature matching via ABC}
\label{sec:gompertz:abc}

Whereas synthetic likelihood carries out many simulations for each likelihood estimation, ABC (as described in \cref{sec:abc}) uses only one.
Each iteration of ABC is therefore much quicker, essentially corresponding to the cost of SMC with a single particle or synthetic likelihood with a single simulation.
A consequence of this is that ABC cannot determine a good relative scaling of the features within each likelihood evaluation and this must be supplied in advance.
One can imagine an adaptive version of ABC which modifies the scaling during the course of the algorithm, but here we do a preliminary calculation to accomplish this.
We return to the Gompertz model to faciliate comparison between ABC and PMCMC.

\begin{knitrout}
\definecolor{shadecolor}{rgb}{1, 1, 1}\color{fgcolor}\begin{kframe}
\begin{verbatim}
R> plist <- list(probe.mean(var = "Y", transform = sqrt),
+                probe.acf("Y", lags = c(0, 5, 10, 20)),
+                probe.marginal("Y", ref = obs(gompertz)))
+  psim <- probe(gompertz, probes = plist, nsim = 500)
+  scale.dat <- apply(psim$simvals, 2, sd)
R> abc1 <- foreach(i = 1:5, .combine = c) %dopar% {
+    abc(pomp(gompertz, dprior = gompertz.dprior), Nabc = 4e6,
+        probes = plist, epsilon = 2, scale = scale.dat,
+        proposal=mvn.diag.rw(c(r = 0.01, sigma = 0.01, tau = 0.01)))
+  }
\end{verbatim}
\end{kframe}
\end{knitrout}

\begin{figure}
\begin{center}
\begin{knitrout}
\definecolor{shadecolor}{rgb}{1, 1, 1}\color{fgcolor}

\includegraphics[width=\maxwidth]{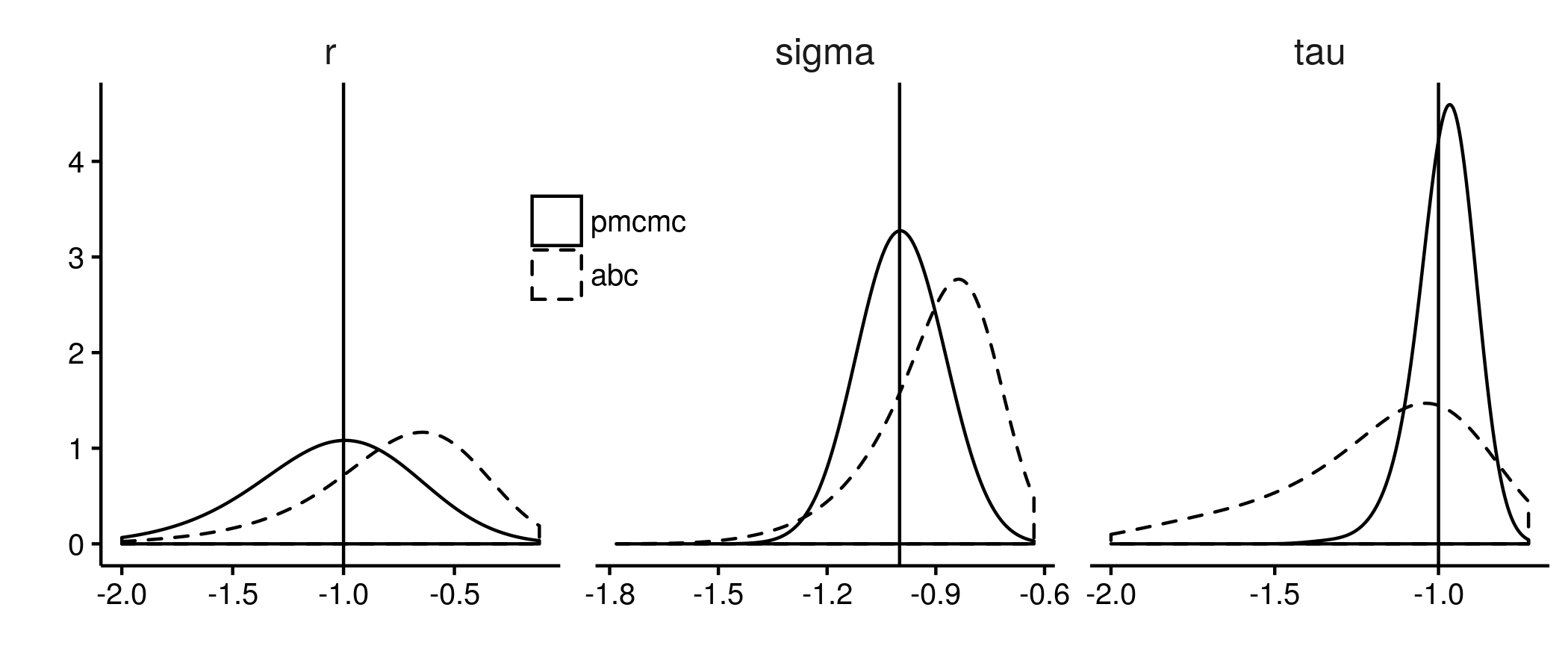} \hfill{}

\end{knitrout}
\end{center}
\caption{
  Marginal posterior distributions using full information via \code{pmcmc} (solid line) and partial information via \code{abc} (dashed line).
  Kernel density estimates are shown for the posterior marginal densities of $\log_{10}({r})$ (left panel), $\log_{10}({\sigma})$ (middle panel), and $\log_{10}({\tau})$ (right panel).
  The vertical lines indicate the true values of each parameter.
}
\label{fig:abc:pmcmc:compare}
\end{figure}

The effective sample size of the ABC chains is lowest for the $r$ parameter (as was the case for PMCMC) and is 450, as compared to 250 for \code{pmcmc} in \cref{sec:gompertz:pmcmc}.
The total computational effort allocated to \code{abc} here matches that for \code{pmcmc} since \code{pmcmc} used 100 particles for each likelihood evaluation but is awarded 100 times fewer Metropolis-Hastings steps.
In this example, we conclude that \code{abc} mixes somewhat more rapidly (as measured by total computational effort) than \code{pmcmc}.
\Cref{fig:abc:pmcmc:compare} investigates the statistical efficiency of \code{abc} on this example.
We see that \code{abc} gives rise to somewhat broader posterior distributions than the full-information posteriors from \code{pmcmc}
As in all numerical studies of this kind, one cannot readily generalize from one particular example: even for this specific model and dataset, the conclusions might be sensitive to the algorithmic settings.
However, one should be aware of the possibility of losing substantial amounts of information even when the features are based on reasoned scientific argument \citep{shrestha11,ionides11-statSci}.
Despite this loss of statistical efficiency, points~B\ref{whyFeature1}--B\ref{whyFeature5} of \cref{sec:probe} identify situations in which ABC may be the only practical method available for Bayesian inference.

\subsection{Parameter estimation by simulated quasi-likelihood}
\label{sec:gompertz:nlf}

Within the \pkg{pomp} environment, it is fairly easy to try a quick comparison to see how \code{nlf} (\cref{sec:nlf}) compares with \code{mif} (\cref{sec:mif}) on the Gompertz model.
Carrying out a simulation study with a correctly specified POMP model is appropriate for assessing computational and statistical efficiency, but does not contribute to the debate on the role of two-step prediction criteria to fit misspecified models \citep{xia11,ionides11-statSci}.
The \code{nlf} implementation we will use to compare to the \code{mif} call from \cref{sec:gompertz:mif} is
\begin{knitrout}
\definecolor{shadecolor}{rgb}{1, 1, 1}\color{fgcolor}\begin{kframe}
\begin{verbatim}
R> nlf1 <- nlf(gompertz, nasymp = 1000, nconverge = 1000, lags = c(2, 3),
+              start = c(r = 1, K = 2, sigma = 0.5, tau = 0.5, X.0 = 1),
+              est = c("r", "sigma", "tau"), transform = TRUE)
\end{verbatim}
\end{kframe}
\end{knitrout}
where the first argument is the \class{pomp} object,
\code{start} is a vector containing model parameters at which \code{nlf}'s search will begin,
\code{est} contains the names of parameters \code{nlf} will estimate, and
\code{lags} specifies which past values are to be used in the autoregressive model.
The \code{transform = TRUE} setting causes the optimization to be performed on the transformed scale, as in \cref{sec:gompertz:mif}.
In the call above \code{lags = c(2, 3)} specifies that the autoregressive model predicts each observation, $y_t$ using $y_{t-2}$ and $y_{t-3}$, as recommended by \citet{kendall05}.
The quasi-likelihood is optimized numerically, so the reliability of the optimization should be assessed by doing multiple fits with different starting parameter values:
the results of a small experiment (not shown) indicate that, on these simulated data, repeated optimization is not needed.
\code{nlf} defaults to optimization by the subplex method \citep{Rowan1990,subplex}, though all optimization methods provided by \code{optim} are available as well.
\code{nasymp} sets the length of the simulation on which the quasi-likelihood is based;
larger values will give less variable parameter estimates, but will slow down the fitting process.
The computational demand of \code{nlf} is dominated by the time required to generate the model simulations, so efficient coding of \code{rprocess} is worthwhile.

\begin{figure}
  \begin{center}
\begin{knitrout}
\definecolor{shadecolor}{rgb}{1, 1, 1}\color{fgcolor}

\includegraphics[width=\maxwidth]{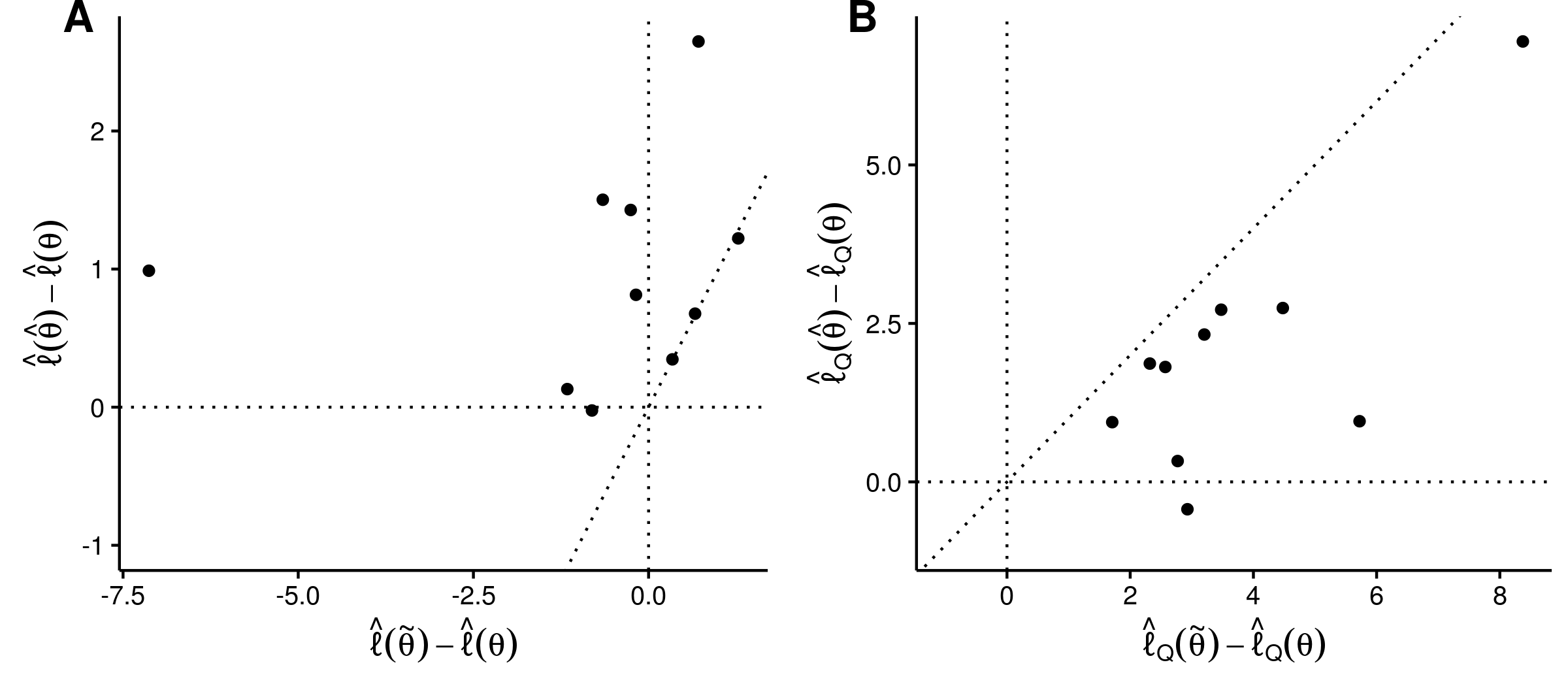} \hfill{}

\end{knitrout}
  \end{center}
  \caption{
    Comparison of \code{mif} and \code{nlf} for 10 simulated datasets using two criteria.
    In both plots, the maximum likelihood estimate (MLE), $\hat\theta$, obtained using iterated filtering is compared with the maximum simulated quasi-likelihood (MSQL) estimate, $\tilde\theta$, obtained using nonlinear forecasting.
    (A) Improvement in estimated log likelihood, $\loglikMC$, at point estimate over that at the true parameter value, $\theta$.
    (B) Improvement in simulated log quasi-likelihood $\loglikMC_Q$, at point estimate over that at the true parameter value, $\theta$.
    In both panels, the diagonal line is the 1-1 line.
    \label{fig:mif:nlf:compare}
  }
\end{figure}

\Cref{fig:mif:nlf:compare} compares the true parameter, $\theta$, with the maximum likelihood estimate (MLE), $\hat\theta$, from \code{mif} and the maximized simulated quasi-likelihood (MSQL), $\tilde\theta$, from \code{nlf}.
\Cref{fig:mif:nlf:compare}A plots  $\loglikMC(\hat\theta)-\loglikMC(\theta)$ against $\loglikMC(\tilde\theta)-\loglikMC(\theta)$, showing that the MSQL estimate can fall many units of log likelihood short of the MLE.
\Cref{fig:mif:nlf:compare}B plots $\loglikMC_Q(\hat\theta)-\loglikMC_Q(\theta)$ against $\loglikMC_Q(\tilde\theta)-\loglikMC_Q(\theta)$, showing that likelihood-based inference is almost as good as \code{nlf} at optimizing the simulated quasi-likelihood criterion which \code{nlf} targets.
\Cref{fig:mif:nlf:compare} suggests that the MSQL may be inefficient, since it can give estimates with poor behavior according to the statistically efficient criterion of likelihood.
Another possibility is that this particular implementation of \code{nlf} was unfortunate.
Each \code{mif} optimization took 26.9~sec to run, compared to 3.7~sec for \code{nlf}, and it is possible that extra computer time or other algorithmic adjustments could substantially improve either or both estimators.
It is hard to ensure a fair comparison between methods, and in practice there is little substitute for some experimentation with different methods and algorithmic settings on a problem of interest.
If the motivation for using NLF is preference for 2-step prediction in place of the likelihood, a comparison with SMC-based likelihood evaluation and maximization is useful to inform the user of the consequences of that preference.

\section{A more complex example: Epidemics in continuous time}
\label{sec:EpidemicModel}

\begin{figure}
  \begin{center}
    \resizebox{0.6\textwidth}{!}{
      \Large
      \setlength{\unitlength}{5pt}
	\begin{picture}(44,20)(-5,-4)
	  \thicklines
	  \put(0,0){\framebox(6,6){$S$}}
	  \put(16,0){\framebox(6,6){$I$}}
	  \put(32,0){\framebox(6,6){$R$}}
	  \put(-4,3){\vector(1,0){4}}
	  \put(-3.7,3.9){$\mu$}
	  \put(6,3){\vector(1,0){10}}
	  \put(9,0){$\lambda(t)$}
	  \put(11,4){\vector(0,1){5}}
	  \put(11.7,6){$\rho$}
	  \put(11,11.5){\circle{5}}
	  \put(9.8,10.6){{$C$}}
	  \put(3,-0.2){\vector(0,-1){4}}
	  \put(4,-3){$\mu$}
	  \put(19,-0.2){\vector(0,-1){4}}
	  \put(20,-3){$\mu$}
	  \put(22,3){\vector(1,0){10}}
	  \put(26,3.9){$\gamma$}
	  \put(35,-0.2){\vector(0,-1){4}}
	  \put(36,-3){$\mu$}
	\end{picture}
    }
  \end{center}
  \caption{
    Diagram of the SIR epidemic model.
    The host population is divided into three classes according to infection status:
    S, susceptible hosts;
    I, infected (and infectious) hosts;
    R, recovered and immune hosts.
    Births result in new susceptibles and all individuals have a common death rate $\mu$.
    Since the birth rate equals the death rate, the expected population size, $P=S+I+R$, remains constant.
    The S$\to$I rate, $\lambda$, called the \emph{force of infection}, depends on the number of infectious individuals according to $\lambda(t)={\beta\,I}/{N}$.
    The I$\to$R, or recovery, rate is $\gamma$.
    The case reports, $C$, result from a process by which new infections are recorded with probability $\rho$.
    Since diagnosed cases are treated with bed-rest and hence removed, infections are counted upon transition to R.
  }
  \label{fig:SIR}
\end{figure}
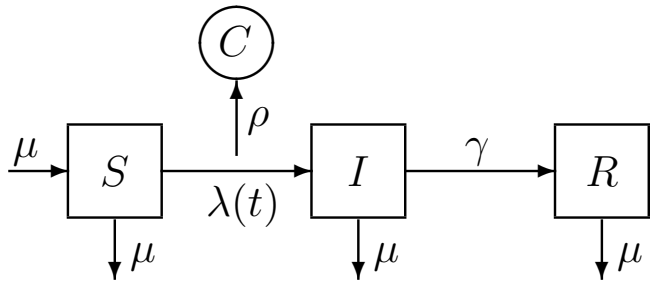

\subsection{A stochastic, seasonal SIR model}

A mainstay of theoretical epidemiology, the SIR model describes the progress of a contagious, immunizing infection through a population of hosts \citep{Kermack1927,Anderson1991,Keeling2008a}.
The hosts are divided into three classes, according to their status vis-{\`a}-vis the infection (\cref{fig:SIR}).
The susceptible class (S) contains those that have not yet been infected and are thereby still susceptible to it;
the infected class (I) comprises those who are currently infected and, by assumption, infectious;
the removed class (R) includes those who are recovered or quarantined as a result of the infection.
Individuals in R are assumed to be immune against reinfection.
We let $S(t)$, $I(t)$, and $R(t)$ represent the numbers of individuals within the respective classes at time $t$.

It is natural to formulate this model as a continuous-time Markov process.
In this process, the numbers of individuals within each class change through time in whole-number increments as discrete births, deaths, and passages between compartments occur.
Let $\cp{A}{B}$ be the stochastic counting process whose value at time $t$ is the number of individuals that have passed from compartment $A$ to compartment $B$ during the interval $[t_0,t)$, where $t_0$ is an arbitrary starting point not later than the first observation.
We use the notation $\cp{\BirthDeath}{S}$ to refer to births and $\cp{A}{\BirthDeath}$ to refer to deaths from compartment A.
Let us assume that the \emph{per capita} birth and death rates, and the rate of transition, $\gamma$, from I to R are constants.
The S to I transition rate, the so-called \emph{force of infection}, $\lambda(t)$, however, should be an increasing function of $I(t)$.
For many infections, it is reasonable to assume that the $\lambda(t)$ is jointly proportional to the fraction of the population infected and the rate at which an individual comes into contact with others.
Here, we will make these assumptions, writing $\lambda(t)=\beta\,I(t)/P$,
where $\beta$ is the transmission rate and $P=S+I+R$ is the population size.
We will go further and assume that birth and death rates are equal and independent of infection status; we will let $\mu$ denote the common rate.
A consequence is that the expected population size remains constant.

The continuous-time Markov process is fully specified by the infinitesimal increment probabilities.
Specifically, writing $\dlta{N(t)}=N(t+h)-N(t)$, we have
\begin{equation}\label{eq:sir-cp}
  \begin{aligned}
    \prob{\dlta{\cp{\BirthDeath}{S}(t)}\myequals 1 \given S(t), I(t), R(t)} &= \mu\,P(t)\,h+o(h), \\
    \prob{\dlta{\cp{S}{I}(t)}\myequals 1 \given S(t), I(t), R(t)} &= \lambda(t)\,S(t)\,h+o(h), \\
    \prob{\dlta{\cp{I}{R}(t)}\myequals 1 \given S(t), I(t), R(t)} &= \gamma\,I(t)\,h+o(h), \\
    \prob{\dlta{\cp{S}{\BirthDeath}(t)}\myequals 1 \given S(t), I(t), R(t) } &= \mu\,S(t)\,h+o(h), \\
    \prob{\dlta{\cp{I}{\BirthDeath}(t)}\myequals 1\given S(t), I(t), R(t)} &= \mu\,I(t)\,h+o(h), \\
    \prob{\dlta{\cp{R}{\BirthDeath}(t)}\myequals 1 \given S(t), I(t), R(t)} &= \mu\,R(t)\,h+o(h),
  \end{aligned}
\end{equation}
together with statement that all events of the form
\begin{equation*}
  \{\dlta{\cp{A}{B}(t)}\,{>}\, 1\} \qquad \text{and} \qquad \{\dlta{\cp{A}{B}(t)}\myequals 1,\dlta{\cp{C}{D}(t)}\myequals 1\}
\end{equation*}
for $A$, $B$, $C$, $D$ with $(A,B)\neq (C,D)$ have probability $o(h)$.
The counting processes are coupled to the state variables \citep{breto11} via the following identities
\begin{equation}\label{eq:sir-cp-bal}
  \begin{aligned}
    \dlta{S}(t) &= \dlta{\cp{\BirthDeath}{S}}(t)-\dlta{\cp{S}{I}}(t)-\dlta{\cp{S}{\BirthDeath}}(t),\\
    \dlta{I}(t) &= \dlta{\cp{S}{I}}(t)-\dlta{\cp{I}{R}}(t)-\dlta{\cp{I}{\BirthDeath}}(t),\\
    \dlta{R}(t) &= \dlta{\cp{I}{R}}(t)-\dlta{\cp{R}{\BirthDeath}}(t).\\
  \end{aligned}
\end{equation}
Taking expectations of \cref{eq:sir-cp,eq:sir-cp-bal}, dividing through by $h$, and taking a limit as $h\downarrow~0$, one obtains a system of nonlinear ordinary differential equations which is known as the deterministic skeleton of the model \citep{coulson04}.
Specifically, the SIR deterministic skeleton is
\begin{equation}
  \begin{aligned}
    &\frac{dS}{dt}=\mu\,(P-S)-\beta\,\frac{I}{P}\,S\\
    &\frac{dI}{dt}=\beta\,\frac{I}{P}\,S-\gamma\,I-\mu\,I\\
    &\frac{dR}{dt}=\gamma\,I-\mu\,R\\
  \end{aligned}
\end{equation}

It is typically impossible to monitor $S$, $I$, and $R$, directly.
It sometimes happens, however, that public health authorities keep records of \emph{cases}, i.e., individual infections.
The number of cases, $C(t_1,t_2)$, recorded within a given reporting interval $[t_1,t_2)$ might perhaps be modeled by a negative binomial process
\begin{equation}
  C(t_1,t_2)\;\sim\;\mathrm{NegBin}(\rho\,\dlta{\cp{S}{I}}(t_1,t_2),\theta)
\end{equation}
where $\dlta{\cp{S}{I}}(t_1,t_2)$ is the true incidence (the accumulated number of new infections that have occured over the $[t_1,t_2)$ interval), $\rho$ is the \emph{reporting rate}, (the probability that an infection is observed and recorded), $\theta$ is the negative binomial ``size'' parameter, and the notation is meant to indicate that $\expect{C(t_1,t_2)\given \dlta{\cp{S}{I}}(t_1,t_2)=H}=\rho\,H$ and $\var{C(t_1,t_2)\given \dlta{\cp{S}{I}}(t_1,t_2)=H}=\rho\,H+\rho^2\,H^2/\theta$.
The fact that the observed data are linked to an accumulation, as opposed to an instantaneous value, introduces a slight complication, which we discuss below.

\subsection[Implementing the SIR model in pomp]{Implementing the SIR model in \pkg{pomp}}

As before, we will need to write functions to implement some or all of the SIR model's \code{rprocess}, \code{rmeasure}, and \code{dmeasure} components.
As in \cref{sec:ricker:setup}, we will write these components using \pkg{pomp}'s \code{Csnippet}s.
Recall that these are snippets of \proglang{C} code that \pkg{pomp} automatically assembles, compiles, and dynamically loads into the running \proglang{R} session.

To start with, we will write snippets that specify the measurement model (\code{rmeasure} and \code{dmeasure}):
\begin{knitrout}
\definecolor{shadecolor}{rgb}{1, 1, 1}\color{fgcolor}\begin{kframe}
\begin{verbatim}
R> rmeas <- "
+    cases = rnbinom_mu(theta, rho * H);
+  "
R> dmeas <- "
+    lik = dnbinom_mu(cases, theta, rho * H, give_log);
+  "
\end{verbatim}
\end{kframe}
\end{knitrout}
Here, we are using \code{cases} to refer to the data (number of reported cases) and \code{H} to refer to the true incidence over the reporting interval.
The negative binomial simulator \code{rnbinom\_mu} and density function \code{dnbinom\_mu} are provided by \pkg{R}.
The logical flag \code{give\_log} requests the likelihood when \code{FALSE}, the log likelihood when \code{TRUE}.
Notice that, in these snippets, we never declare the variables;
\code{pomp} will ensure that the state variable (\code{H}), observable (\code{cases}), parameters (\code{theta}, \code{rho}), and likelihood (\code{lik}) are defined in the contexts within which these snippets are executed.

For the \code{rprocess} portion, we could simulate from the continuous-time Markov process exactly \citep{gillespie77};
the \pkg{pomp} function \code{gillespie.sim} implements this algorithm.
However, for practical purposes, the exact algorithm is often prohibitively slow.
If we are willing to live with an approximate simulation scheme, we can use the so-called ``tau-leap'' algorithm, one version of which is implemented in \pkg{pomp} via the \code{euler.sim} plug-in.
This algorithm holds the transition rates $\lambda$, $\mu$, $\gamma$ constant over a small interval of time $\dlta{t}$ and simulates the numbers of births, deaths, and transitions that occur over that interval.
It then updates the state variables $S$, $I$, $R$ accordingly, increments the time variable by $\dlta{t}$, recomputes the transition rates, and repeats.
Naturally, as $\dlta{t}\to 0$, this approximation to the true continuous-time process becomes better and better.
The critical feature from the inference point of view, however, is that no relationship need be assumed between the Euler simulation interval $\dlta{t}$ and the reporting interval, which itself need not even be the same from one observation to the next.

Under the above assumptions, the number of individuals leaving any of the classes by all available routes over a particular time interval is a multinomial process.
For example, if $\dlta{\cp{S}{I}}$ and $\dlta{\cp{S}{}}$ are the numbers of S individuals acquiring infection and dying, respectively, over the Euler simulation interval $[t,t+\dlta{t})$, then
\begin{equation}\label{eq:eulermultinomial}
    (\dlta{\cp{S}{I}},\dlta{\cp{S}{}},S-\dlta{\cp{S}{I}}-\dlta{\cp{S}{}})\sim\mathrm{Multinom}\left(S(t);p_{S{\to}I},p_{S{\to}},1-p_{S{\to}I}-p_{S{\to}}\right),\\
\end{equation}
where
\begin{equation}\label{eq:eulermultinomial2}
  \begin{aligned}
    p_{S{\to}I} &= \frac{\lambda(t)}{\lambda(t)+\mu}\,\left(1-e^{-(\lambda(t)+\mu)\,\dlta{t}}\right)\\
    p_{S{\to}} &= \frac{\mu}{\lambda(t)+\mu}\,\left(1-e^{-(\lambda(t)+\mu)\,\dlta{t}}\right).
  \end{aligned}
\end{equation}
By way of shorthand, we say that the random variable $(\dlta{\cp{S}{I}},\dlta{\cp{S}{}})$ in \cref{eq:eulermultinomial} has an \emph{Euler-multinomial} distribution.
Such distributions arise with sufficient frequency in compartmental models that \pkg{pomp} provides convenience functions for them.
Specifically, the functions \code{reulermultinom} and \code{deulermultinom} respectively draw random deviates from, and evaluate the probability mass function of, such distributions.
As the help pages relate, \code{reulermultinom} and \code{deulermultinom} parameterize the Euler-multinomial distributions by the size ($S(t)$ in \cref{eq:eulermultinomial}), rates ($\lambda(t)$ and $\mu$), and time interval $\dlta{t}$.
Obviously, the Euler-multinomial distributions generalize to an arbitrary number of exit routes.

The help page (\code{?euler.sim}) informs us that to use \code{euler.sim}, we need to specify a function that advances the states from $t$ to $t+\dlta{t}$.
Again, we will write this in \proglang{C} to realize faster run-times:
\begin{knitrout}
\definecolor{shadecolor}{rgb}{1, 1, 1}\color{fgcolor}\begin{kframe}
\begin{verbatim}
R> sir.step <- "
+    double rate[6];
+    double dN[6];
+    double P;
+    P = S + I + R;
+    rate[0] = mu * P;       // birth
+    rate[1] = beta * I / P; // transmission
+    rate[2] = mu;           // death from S
+    rate[3] = gamma;        // recovery
+    rate[4] = mu;           // death from I
+    rate[5] = mu;           // death from R
+    dN[0] = rpois(rate[0] * dt);
+    reulermultinom(2, S, &rate[1], dt, &dN[1]);
+    reulermultinom(2, I, &rate[3], dt, &dN[3]);
+    reulermultinom(1, R, &rate[5], dt, &dN[5]);
+    S += dN[0] - dN[1] - dN[2];
+    I += dN[1] - dN[3] - dN[4];
+    R += dN[3] - dN[5];
+    H += dN[1];
+  "
\end{verbatim}
\end{kframe}
\end{knitrout}

As before, \pkg{pomp} will ensure that the undeclared state variables and parameters are defined in the context within which the snippet is executed.
Note, however, that in the above we do declare certain local variables.
In particular, the \code{rate} and \code{dN} arrays hold the rates and numbers of transition events, respectively.
Note too, that we make use of \pkg{pomp}'s \proglang{C} interface to \code{reulermultinom}, documented in the package help pages (\code{?reulermultinom}).
The package help system (\code{?Csnippet}) includes instructions for, and examples of, the use of \code{Csnippet}s.

Two significant wrinkles remains to be explained.
First, notice that in \code{sir.step}, the variable \code{H} simply accumulates the numbers of new infections: \code{H} is a counting process that is nondecreasing with time.
In fact, the incidence within an interval $[t_1,t_2)$ is $\dlta{\cp{S}{I}}(t_1,t_2)=\mathtt{H}(t_2)-\mathtt{H}(t_1)$.
This leads to a technical difficulty with the measurement process, however, in that the data are assumed to be records of new infections occurring within the latest reporting interval, while the process model tracks the accumulated number of new infections since time $t_0$.
We can get around this difficulty by re-setting \code{H} to zero immediately after each observation.
We cause \pkg{pomp} to do this via the \code{pomp} function's \code{zeronames} argument, as we will see in a moment.
The section on ``accumulator variables'' in the \code{pomp} help page (\code{?pomp}) discusses this in more detail.

The second wrinkle has to do with the initial conditions, i.e., the states $S(t_0)$, $I(t_0)$, $R(t_0)$.
By default, \pkg{pomp} will allow us to specify these initial states arbitrarily.
For the model to be consistent, they should be positive integers that sum to the population size $N$.
We can enforce this constraint by customizing the parameterization of our initial conditions.
We do this in by furnishing a custom \code{initializer} in the call to \code{pomp}.
Let us construct it now and fill it with simulated data.
\begin{knitrout}
\definecolor{shadecolor}{rgb}{1, 1, 1}\color{fgcolor}\begin{kframe}
\begin{verbatim}
R> pomp(data = data.frame(cases = NA, time = seq(0, 10, by=1/52)),
+       times = "time", t0 = -1/52, dmeasure = Csnippet(dmeas),
+       rmeasure = Csnippet(rmeas), rprocess = euler.sim(
+         step.fun = Csnippet(sir.step), delta.t = 1/52/20),
+       statenames = c("S", "I", "R", "H"),
+       paramnames = c("gamma", "mu", "theta", "beta", "popsize",
+         "rho", "S.0", "I.0", "R.0"), zeronames=c("H"),
+       initializer=function(params, t0, ...) {
+         fracs <- params[c("S.0", "I.0", "R.0")]
+         setNames(c(round(params["popsize"]*fracs/sum(fracs)),0),
+                  c("S","I","R","H"))
+       }, params = c(popsize = 500000, beta = 400, gamma = 26,
+            mu = 1/50, rho = 0.1, theta = 100, S.0 = 26/400,
+            I.0 = 0.002, R.0 = 1)) -> sir1
R> simulate(sir1, seed = 1914679908L) -> sir1
\end{verbatim}
\end{kframe}
\end{knitrout}

\begin{figure}
\begin{knitrout}
\definecolor{shadecolor}{rgb}{1, 1, 1}\color{fgcolor}

\includegraphics[width=\maxwidth]{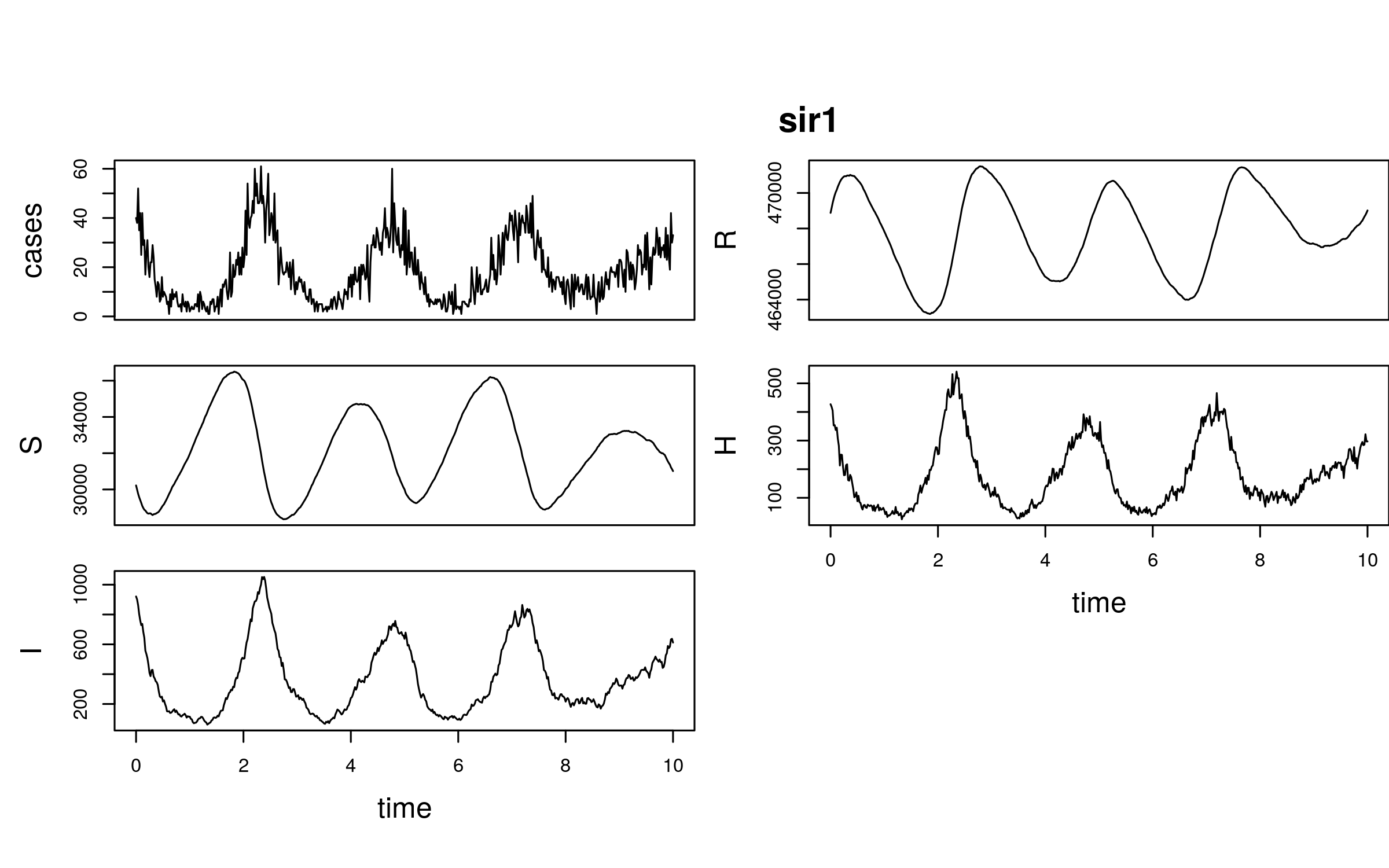} \hfill{}

\end{knitrout}
  \caption{
    Result of \code{plot(sir1)}.
    The \class{pomp} object \code{sir1} contains the SIR model with simulated data.
  }
  \label{fig:sir1-plot}
\end{figure}

Notice that we are assuming here that the data are collected weekly and use an Euler step-size of 1/20~wk.
Here, we have assumed an infectious period of 2~wk ($1/\gamma=1/26$~yr) and a basic reproductive number, $R_0$ of $\beta/(\gamma+\mu)\approx 15$.
We have assumed a host population size of 500,000 and 10\% reporting efficiency.
\cref{fig:sir1-plot} shows one realization of this process.

\clearpage
\subsection{Complications: Seasonality, imported infections, extra-demographic stochasticity}

To illustrate the flexibility afforded by \pkg{pomp}'s plug-and-play methods, let us add a bit of real-world complexity to the simple SIR model.
We will modify the model to take four facts into account:
\begin{enumerate}
  \item For many infections, the contact rate is \emph{seasonal}: $\beta=\beta(t)$ varies in more or less periodic fashion with time.
  \item The host population may not be truly closed: \emph{imported infections} arise when infected individuals visit the host population and transmit.
  \item The host population need not be constant in size.
    If we have data, for example, on the numbers of births occurring in the population, we can incorporate this directly into the model.
  \item Stochastic fluctuation in the rates $\lambda$, $\mu$, and $\gamma$ can give rise to \emph{extrademographic stochasticity}, i.e., random process variability beyond the purely demographic stochasticity we have included so far.
\end{enumerate}

To incorporate seasonality, we would like to assume a flexible functional form for $\beta(t)$.
Here, we will use a three-coefficient Fourier series:
\begin{equation}
  \log{\beta(t)}=b_0+b_1\,\cos{2\pi t}+b_2\sin{2\pi t}.
\end{equation}

There are a variety of ways to account for imported infections.
Here, we will simply assume that there is some constant number, $\iota$, of infected hosts visiting the population.
Putting this together with the seasonal contact rate results in a force of infection $\lambda(t)=\beta(t)\,\left(I(t)+\iota\right)/N$.

To incorporate birth-rate information, let us suppose we have data on the number of births occurring each month in this population and that these data are in the form of a data frame \code{birthdat} with columns \code{time} and \code{births}.
We can incorporate the varying birth rate into our model by passing it as a covariate to the simulation code.
When we pass \code{birthdat} as the \code{covar} argument to \code{pomp}, we cause a look-up table to be created and made available to the simulator.
The package employs linear interpolation to provide a value of each variable in the covariate table at any requisite time:
from the user's perspective, a variable \code{births} will simply be available for use by the model codes.

Finally, we can allow for extrademographic stochasticity by allowing the force of infection to be itself a random variable.
We will accomplish this by assuming a random phase in $\beta$:
\begin{equation}
  \lambda(t) = \left(\beta(\Phi(t))\,\frac{I(t)+\iota}{N}\right)
\end{equation}
where the phase $\Phi$ satisfies the stochastic differential equation
\begin{equation}
  d\Phi=dt+\sigma\,dW_t,
\end{equation}
where $dW(t)$ is a white noise, specifically an increment of standard Brownian motion.
This model assumption attempts to capture variability in the timing of seasonal changes in transmission rates.
As $\sigma$ varies, it can represent anything from a very mild modulation of the timing of the seasonal progression to much more intense variation.

Let us modify the process-model simulator to incorporate these complexities.
\begin{knitrout}
\definecolor{shadecolor}{rgb}{1, 1, 1}\color{fgcolor}\begin{kframe}
\begin{verbatim}
R> seas.sir.step <- "
+    double rate[6];
+    double dN[6];
+    double Beta;
+    double dW;
+    Beta = exp(b1 + b2 * cos(M_2PI * Phi) + b3 * sin(M_2PI * Phi));
+    rate[0] = births;                // birth
+    rate[1] = Beta * (I + iota) / P; // infection
+    rate[2] = mu;                    // death from S
+    rate[3] = gamma;                 // recovery
+    rate[4] = mu;                    // death from I
+    rate[5] = mu;                    // death from R
+    dN[0] = rpois(rate[0] * dt);
+    reulermultinom(2, S, &rate[1], dt, &dN[1]);
+    reulermultinom(2, I, &rate[3], dt, &dN[3]);
+    reulermultinom(1, R, &rate[5], dt, &dN[5]);
+    dW = rnorm(dt, sigma * sqrt(dt));
+    S += dN[0] - dN[1] - dN[2];
+    I += dN[1] - dN[3] - dN[4];
+    R += dN[3] - dN[5];
+    P = S + I + R;
+    Phi += dW;
+    H += dN[1];
+    noise += (dW - dt) / sigma;
+  "
R> pomp(sir1, rprocess = euler.sim(
+    step.fun = Csnippet(seas.sir.step), delta.t = 1/52/20),
+    dmeasure = Csnippet(dmeas), rmeasure = Csnippet(rmeas),
+    covar = birthdat, tcovar = "time", zeronames = c("H", "noise"),
+    statenames = c("S", "I", "R", "H", "P", "Phi", "noise"),
+    paramnames = c("gamma", "mu", "popsize", "rho","theta","sigma",
+                   "S.0", "I.0", "R.0", "b1", "b2", "b3", "iota"),
+    initializer = function(params, t0, ...) {
+      fracs <- params[c("S.0", "I.0", "R.0")]
+      setNames(c(round(params["popsize"]*c(fracs/sum(fracs),1)),0,0,0),
+               c("S","I","R","P","H","Phi","noise"))
+    }, params = c(popsize = 500000, iota = 5, b1 = 6, b2 = 0.2,
+                  b3 = -0.1, gamma = 26, mu = 1/50, rho = 0.1, theta = 100,
+                  sigma = 0.3, S.0 = 0.055, I.0 = 0.002, R.0 = 0.94)) -> sir2
R> simulate(sir2, seed = 619552910L) -> sir2
\end{verbatim}
\end{kframe}
\end{knitrout}
\Cref{fig:sir2-plot} shows the simulated data and latent states.
The \code{sir2} object we have constructed here contains all the key elements of models used within the \pkg{pomp} to investigate cholera \citep{king08}, measles \citep{he10}, malaria \citep{bhadra11}, pertussis \citep{blackwood13,Lavine2013}, pneumococcal pneumonia \citep{shrestha13}, rabies \citep{blackwood13b}, and Ebola virus disease \citep{King2015}.

\begin{figure}
\begin{knitrout}
\definecolor{shadecolor}{rgb}{1, 1, 1}\color{fgcolor}

\includegraphics[width=\maxwidth]{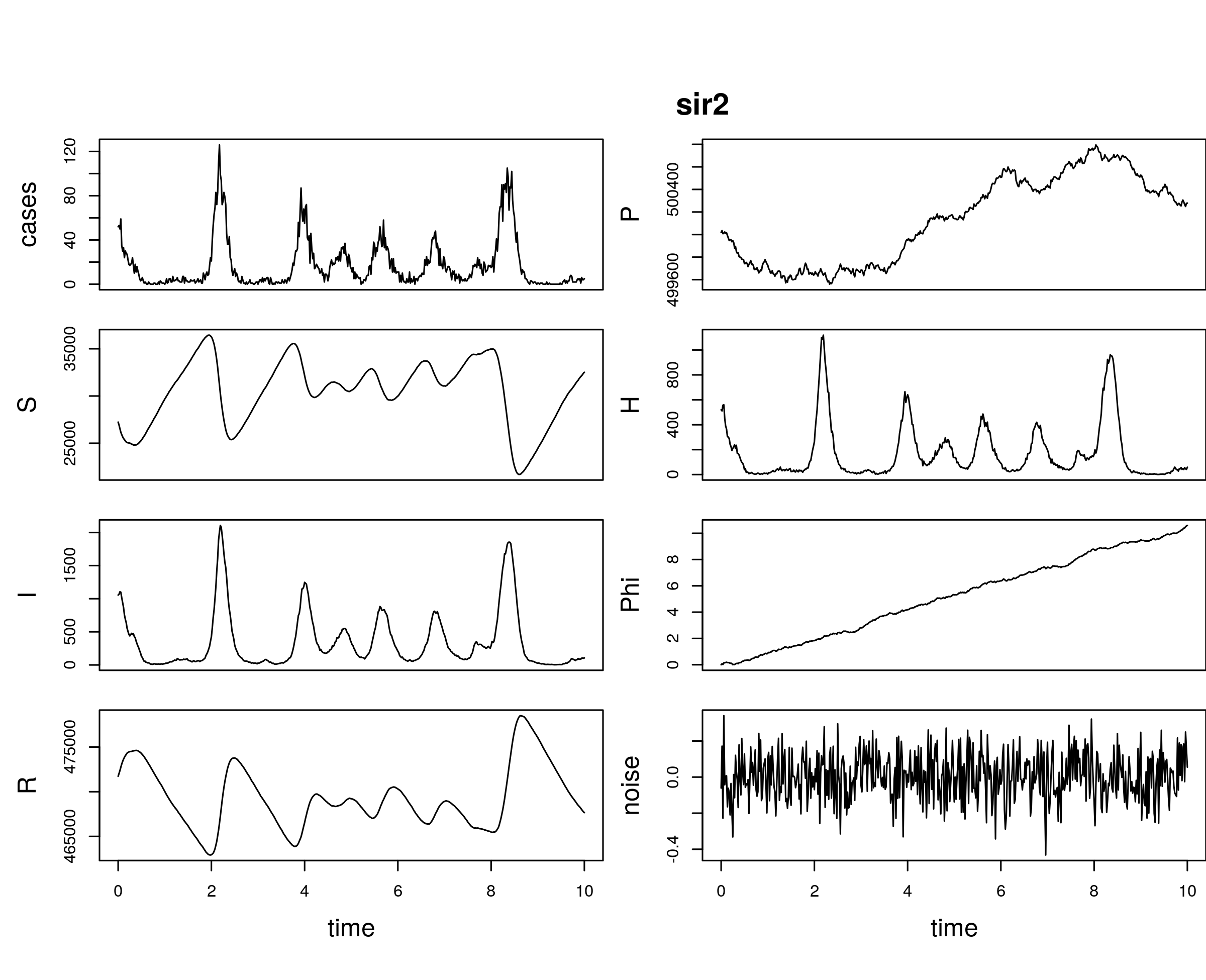} \hfill{}

\end{knitrout}
  \caption{
    One realization of the SIR model with seasonal contact rate, imported infections, and extrademographic stochasticity in the force of infection.
}
  \label{fig:sir2-plot}
\end{figure}

\section{Conclusion}
\label{sec:conclusion}

The \pkg{pomp} package is designed to be both a tool for data analysis based on POMP models and a sound platform for the development of inference algorithms.
The model specification language provided by \pkg{pomp} is very general.
Implementing a POMP model in \pkg{pomp} makes a wide range of inference algorithms available.
Moreover, the separation of model from inference algorithm facilitates objective comparison of alternative models and methods.
The examples demonstrated in this paper are relatively simple, but the package has been instrumental in a number of scientific studies \citep[e.g.,][]{king08,bhadra11,shrestha11,Earn2012,roy13,shrestha13,blackwood13,blackwood13b,Lavine2013,He2013a,Breto2014,King2015}.

As a development platform, \pkg{pomp} is particularly convenient for implementing algorithms with the plug-and-play property, since models will typically be defined by their \code{rprocess} simulator, together with \code{rmeasure} and often \code{dmeasure}, but can accommodate inference methods based on other model components such as \code{dprocess} and \code{skeleton} (the deterministic skeleton of the latent process).
As an open-source project, the package readily supports expansion, and the authors invite community participation in the \pkg{pomp} project in the form of additional inference algorithms, improvements and extensions of existing algorithms, additional model/data examples, documentation contributions and improvements, bug reports, and feature requests.

Complex models and large datasets can challenge computational resources.
With this in mind, key components of the \pkg{pomp} are written in \proglang{C}, and \pkg{pomp} provides facilities for users to write models either in \proglang{R} or, for the acceleration that typically proves necessary in applications, in \proglang{C}.
Multi-processor computing also becomes necessary for ambitious projects.
The two most common computationally intensive tasks are assessment of Monte~Carlo variability and investigation of the role of starting values and other algorithmic settings on optimization routines.
These analyses require only embarrassingly parallel computations and need no special discussion here.

The package contains more examples (via \code{pompExample}), which can be used as templates for implementation of new models;
the \proglang{R} and \proglang{C} code underlying these examples is provided with the package.
In addition, \pkg{pomp} provides a number of interactive demos (via \code{demo}).
Further documentation and an introductory tutorial are provided with the package and on the \pkg{pomp} website, \url{http://kingaa.github.io/pomp}.

\paragraph*{Acknowledgments}

Initial development of \pkg{pomp} was carried out as part of the {\it Inference for Mechanistic Models} working group supported from 2007 to 2010 by the National Center for Ecological Analysis and Synthesis, a center funded by the U.S.\ National Science Foundation (Grant DEB-0553768), the University of California, Santa Barbara, and the State of California.
Participants were C.~Bret\'{o}, S.~P.~Ellner, M.~J.~Ferrari, G.~J.~Gibson, G.~Hooker, E.~L.~Ionides, V.~Isham, B.~E.~Kendall, K.~Koelle, A.~A.~King, M.~L.~Lavine, K.~B.~Newman, D.~C.~Reuman, P.~Rohani and H.~J.~Wearing.
As of this writing, the \pkg{pomp} development team is A.~A.~King, E.~L.~Ionides, and D.~Nguyen.
Financial support was provided by grants DMS-1308919, DMS-0805533, EF-0429588 from the U.S.\ National Science Foundation and by the Research and Policy for Infectious Disease Dynamics program of the Science and Technology Directorate, U.S.\ Department of Homeland Security and the Fogarty International Center, U.S.\ National Institutes of Health.

\bibliography{pompjss}

\end{document}